\documentclass [10pt,twocolumn]{IEEEtran}
\pdfoutput=1
\usepackage{algorithm,algorithmic,amsbsy,amssymb,epsfig,bbm,bm,mathrsfs,multirow,amsthm,listings,xcolor}
\usepackage[T1]{fontenc}
\usepackage{amsmath}
\usepackage{float}
\usepackage{cite}
\usepackage{array,multirow,graphicx}
\usepackage{color}
\usepackage{makecell}
\usepackage[hidelinks]{hyperref}
\usepackage[flushleft]{threeparttable}

\colorlet{punct}{red!60!black}
\definecolor{background}{HTML}{F7F7F7}
\definecolor{delim}{RGB}{20,105,176}
\colorlet{numb}{magenta!60!black}
\lstdefinelanguage{json}{
    basicstyle=\scriptsize\ttfamily,
    numbers=left,
    numberstyle=\scriptsize,
    stepnumber=1,
    numbersep=8pt,
    showstringspaces=false,
    breaklines=true,
    frame=lines,
    backgroundcolor=\color{background},
    literate=
     *{0}{{{\color{numb}0}}}{1}
      {1}{{{\color{numb}1}}}{1}
      {2}{{{\color{numb}2}}}{1}
      {3}{{{\color{numb}3}}}{1}
      {4}{{{\color{numb}4}}}{1}
      {5}{{{\color{numb}5}}}{1}
      {6}{{{\color{numb}6}}}{1}
      {7}{{{\color{numb}7}}}{1}
      {8}{{{\color{numb}8}}}{1}
      {9}{{{\color{numb}9}}}{1}
      {:}{{{\color{punct}{:}}}}{1}
      {,}{{{\color{punct}{,}}}}{1}
      {\{}{{{\color{delim}{\{}}}}{1}
      {\}}{{{\color{delim}{\}}}}}{1}
      {[}{{{\color{delim}{[}}}}{1}
      {]}{{{\color{delim}{]}}}}{1}
      {\ \ }{\ }1,
}

\begin{document}

\title{Enabling Technologies for Web 3.0: \\ A Comprehensive Survey}
\author{\IEEEauthorblockN{Md Arif Hassan\IEEEauthorrefmark{1}, Mohammad (Behdad) Jamshidi\IEEEauthorrefmark{1}, Bui Duc Manh\IEEEauthorrefmark{1}, Nam H. Chu\IEEEauthorrefmark{1}, Chi-Hieu Nguyen\IEEEauthorrefmark{1}, \\ Nguyen Quang Hieu\IEEEauthorrefmark{1}, Cong T. Nguyen\IEEEauthorrefmark{1}, Dinh Thai Hoang\IEEEauthorrefmark{1}, Diep N. Nguyen\IEEEauthorrefmark{1}, Nguyen Van Huynh\IEEEauthorrefmark{2}, Mohammad Abu Alsheikh\IEEEauthorrefmark{3}, and Eryk Dutkiewicz\IEEEauthorrefmark{1}}
	
 \IEEEauthorblockA{\IEEEauthorrefmark{1}School of Electrical and Data Engineering, University of Technology Sydney, Australia} \\
 \IEEEauthorblockA{\IEEEauthorrefmark{2}School of Computing, Engineering, and the Built Environment, Edinburgh Napier University, Scotland} \\
 \IEEEauthorblockA{\IEEEauthorrefmark{3}Faculty of Science and Technology, University of Canberra, Australia}

}
\maketitle
\begin{abstract}
Web 3.0 represents the next stage of Internet evolution, aiming to empower users with increased autonomy, efficiency, quality, security, and privacy. This evolution can potentially democratize content access by utilizing the latest developments in enabling technologies. In this paper, we conduct an in-depth survey of enabling technologies in the context of Web 3.0, such as blockchain, semantic web, 3D interactive web, Metaverse, Virtual reality/Augmented reality, Internet of Things technology, and their roles in shaping Web 3.0. We commence by providing a comprehensive background of Web 3.0, including its concept, basic architecture, potential applications, and industry adoption. Subsequently, we examine recent breakthroughs in IoT, 5G, and blockchain technologies that are pivotal to Web 3.0 development. Following that, other enabling technologies, including AI, semantic web, and 3D interactive web, are discussed. Utilizing these technologies can effectively address the critical challenges in realizing Web 3.0, such as ensuring decentralized identity, platform interoperability, data transparency, reducing latency, and enhancing the system's scalability. Finally, we highlight significant challenges associated with Web 3.0 implementation, emphasizing potential solutions and providing insights into future research directions in this field.
\end{abstract}

\begin{keywords}
Web 3.0, Internet of Things, 5G and beyond, blockchain, semantic web, Metaverse, 3D interactive web, Artificial intelligence, decentralized networks, security and privacy.
\end{keywords}

\section{Introduction}
\label{sec:Intro}

\begin{table*}
	\footnotesize
	\centering
	\caption{\footnotesize List of Abbreviations} \label{tab:abbreviation} 
	\begin{tabular}{|>{\raggedright\arraybackslash}m{1.5cm}|>{\raggedright\arraybackslash}m{3.5cm}|>{\raggedright\arraybackslash}m{1.5cm}|>{\raggedright\arraybackslash}m{3.5cm}|>{\raggedright\arraybackslash}m{1.5cm}|>{\raggedright\arraybackslash}m{3.5cm}|} 
		\hline
		\multicolumn{1}{|>{\centering\arraybackslash}m{1.5cm}|}{\textbf{Abbreviation}} & 
        \multicolumn{1}{>{\centering\arraybackslash}m{2.5cm}|}{\textbf{Description}} & 
        \multicolumn{1}{|>{\centering\arraybackslash}m{1.5cm}|}{\textbf{Abbreviation}} & 
        \multicolumn{1}{>{\centering\arraybackslash}m{3.5cm}|}{\textbf{Description}} & 
        \multicolumn{1}{>{\centering\arraybackslash}m{1.5cm}|}{\textbf{Abbreviation}} & 
        \multicolumn{1}{>{\centering\arraybackslash}m{3.5cm}|}{\textbf{Description}} 
        \\ \hline		\hline

         AR &  Augmented reality & BTC & Bitcoin & DAO & Decentralized Autonomous Organization  \\ \hline

        DApps & Decentralized Application & ETH & Ethereum  &  FL  &  File Inclusion \\ \hline

        Intel SGX & Intel Software Guard Extensions& IoT  & Internet of Things & IPFS & Interplanetary File System \\ \hline

        JSON & JavaScript Object Notation &  ML & Machine Learning & MIMO & Multiple-Input Multiple-Output  \\ \hline

        NFT & Non-fungible Tokens & OWL &  Web Ontology Language & PoS & Proof of Stake \\ \hline

        PoW & Proof-of-Work & QIT & Quantum Information Technology & RDF & Resource Description Framework  \\ \hline

       SMW & Social Media Web & SWT & Semantic Web Technologies  &  Turtle & Terse RDF Triple Language  \\ \hline

       3D &  Three Dimensional & URI & Universal Resource Identifier & VR & Virtual Reality  \\ \hline

       WWW & World Wide Web & WebGL & Web Graphics Library &  W3C & World Wide Web Consortium \\ \hline

	\end{tabular} 
\end{table*}

\PARstart{T}HE World Wide Web (WWW) has undergone significant transformations since its launch in 1989\cite{webfoundation}. Initially, the first generation of WWW (Web 1.0) aimed to provide users with access to static data, e.g., pictures and text on non-interactive web pages. Then, the advent of Web 2.0 in the early 2000s introduced innovative components such as interactive tools and dynamic content to websites. This helped to transform the WWW from a read-only, static presentation of information to an interactive and continuously evolving platform. This transformation enables users to not only read but also create and share content via making blogs, wikis, social media, and online collaborative platforms~\cite{salim2021blockchain}.  As a result, Web 2.0 brings significant benefits, ranging from higher interactivity, easier online collaborations to new business opportunities and the availability of online education.

Despite those significant advantages, Web 2.0 is also facing many technical challenges. First, Web 2.0 still relies on centralized platforms with significant control over users' content, which can be manipulated or monetized without user consent. Moreover, trust and privacy are primary concerns as users' must rely on intermediaries, thereby exposing themselves to cyber security risks. Furthermore, Web 2.0 struggles with limited intelligence and connectivity through keyword-based search and algorithms, often delivering irrelevant or biased results~\cite{korkmaz2022alder}. In practice, Web 2.0 is currently facing serious issues on data privacy breaches, disinformation spreading, and the concentration of power in the hands of big companies, which demand serious consideration and solutions~\cite{lin2016health} \cite{stritter2016cleaning}. Failure to address these issues can cause serious consequences on individuals as well as society in general.

 \begin{figure}
    \includegraphics[width=1.0\linewidth]{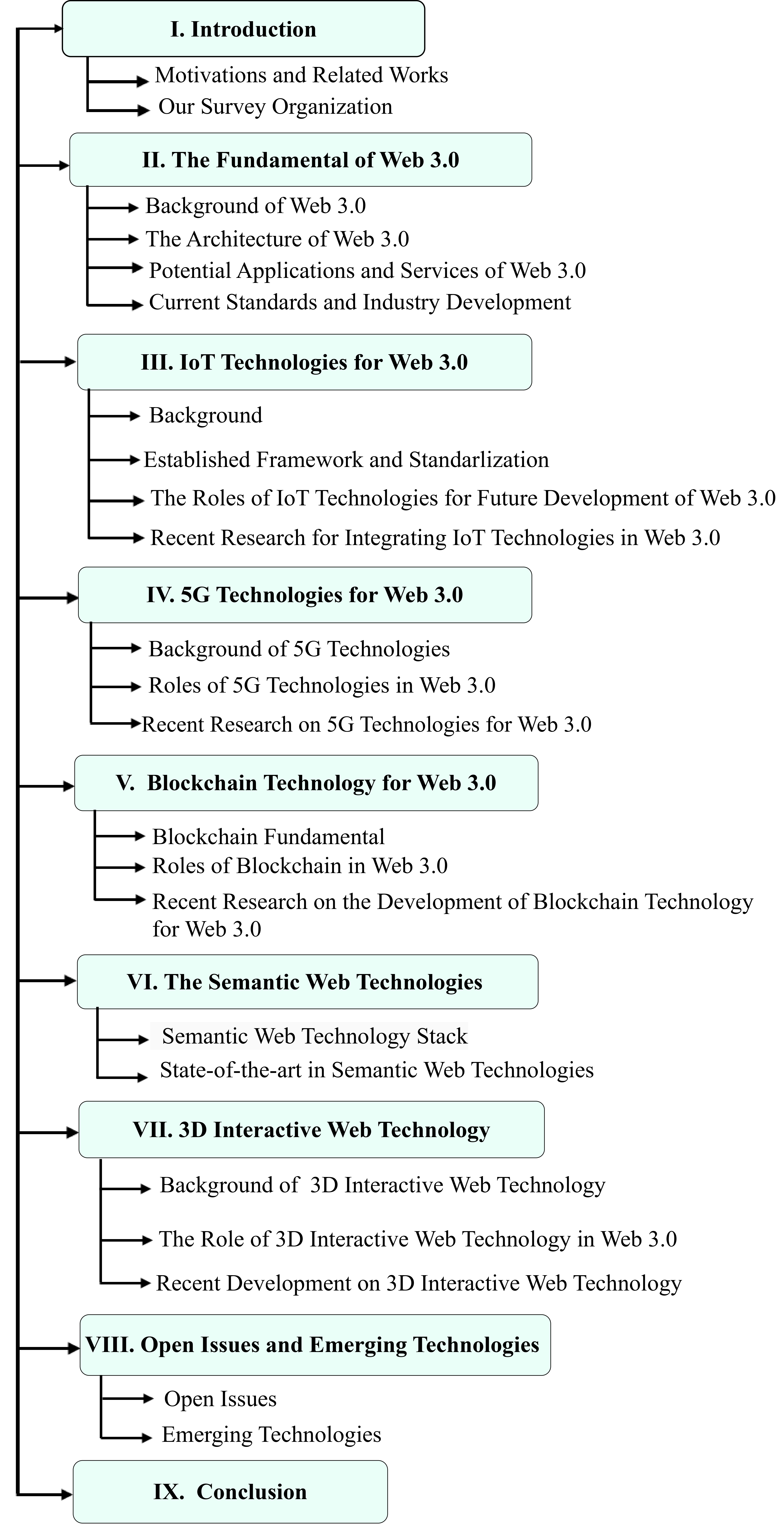}
    \caption{The organization of this paper.}
    \label{fig:overview}
\end{figure}


Web 3.0 has been recently introduced as a breakthrough technology to overcome the limitations of Web 2.0, which is more intelligent, decentralized, and user-centric web. Relying on cutting-edge technologies such as blockchain, Artificial intelligence (AI), Internet of Things (IoT), Semantic web, and 3D interactive web, Web 3.0 aims to improve security and privacy while giving users more authority over their generated contents\cite{liu2023slicing4meta}. For example, blockchain technology plays a pivotal role in enabling decentralized information management and providing users with greater control over their own data. In this way, we can not only address issues related to data ownership and censorship but also pave the way for a number of new services for users and providers, such as decentralized finance (DeFi) and Non-Fungible Tokens (NFTs). Moreover, the integration of AI and IoT can make Web 3.0 more intelligent and responsive to improve personalized experiences toward a highly efficient web platform-based ecosystem. Alternatively, semantic web technology enables to automatically and smartly process data on a semantic level, thereby enhancing the efficiency of information processing. Additionally, semantic technology and 3D web interactive technology can enhance users' experiences by improving Web 3.0 personalized services and immersive visualization. Therefore, Web 3.0 represents a more open, connected, intelligent, and user-empowered Internet. It promises to reshape how we interact with the web, making it more secure, private, and user-centric, while also creating new opportunities for commerce, communication, and content creation.

This paper aims to provide an in-depth and comprehensive survey on enabling technologies for Web 3.0. 
Particularly, we first provide a fundamental background on Web 3.0, including key concepts, basic architecture, potential applications, as well as current standards and industry development. Then, a comprehensive survey on the enabling technologies of Web 3.0 is presented, including IoT, 5G, blockchain, semantic web, and 3D interactive web. For each technology, we first provide a quick overview of the fundamental concepts, discuss the role of technology in Web 3.0, and survey the recent research and development of the technologies for Web 3.0. Then, we discuss the open issues that Web 3.0 is currently facing along with their potential solutions. Finally, promising future research directions are presented.  

There are a few surveys related to Web 3.0 with different focuses. Particularly,~\cite{bashir2023systematic} explores the applications Web 3.0 for e-learning, whereas~\cite{chen2022digital} discusses the digital economy in Web 3.0. Moreover, \cite{zarrin2021blockchain} and \cite{ren2023building} discuss the role of blockchain technology in Web 3.0, and~\cite{ray2023web3} focuses mainly on the applications of Web 3.0. However, to the best of our knowledge, there is currently no comprehensive survey that thoroughly examines the enabling technologies for Web 3.0. Given the explosive growth of attention to the development of Web 3.0 and the indisputable roles of cutting-edge technologies, such as blockchain, 5G, semantic web, 3D interactive web, and IoT, this review paper aims to fill the current gap in the literature by providing a comprehensive overview and facilitating further research, innovation, and adoption of those technologies for the development of Web 3.0.

As shown in Fig.~\ref{fig:overview}, the rest of this article is organized as follows. Section~\ref{sec:Web3.0} covers the fundamentals of Web 3.0, including concepts, architecture, applications, and current industry standards and services. Then, the core enabling technologies, such as blockchain, 5G, semantic web, 3D interactive web, and IoT, are discussed from Section~\ref{sec:iot} to Section~\ref{sec:3D_web}. After that, open issues and emerging technologies for future development are discussed in Section~\ref{sec:Dis}. Finally, Section~\ref{sec:Conclusion} concludes the paper. The abbreviations used in this article are summarized in Table~\ref{tab:abbreviation}.

\begin{figure*}[!]
    \centering
    \includegraphics[width=0.8\linewidth]{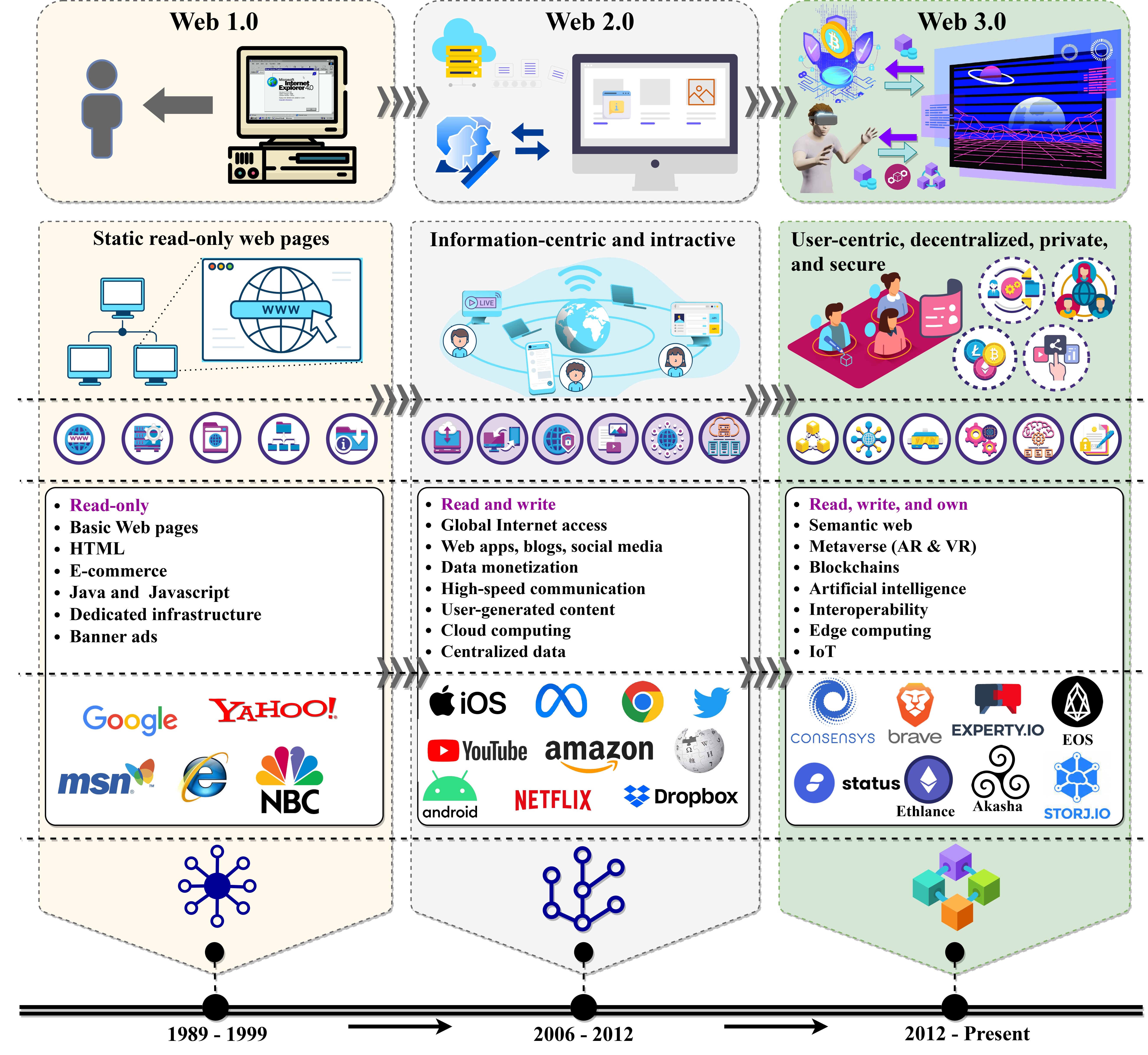}
    \caption{The evaluation of the web for different perspectives.}
    \label{fig:web1toweb3}
\end{figure*}

\section{The Fundamental of Web 3.0}
\label{sec:Web3.0}
\subsection{Background of Web 3.0}
In recent years, we have envisioned the reborn interest of Web 3.0, which was previously named ``Semantic Web" or ``Decentralize Web". Although the concepts of semantic web and decentralized web date back to the early 2000s, technological advancements at that time made it difficult to fully realize Web 3.0. Fortunately, the recently ever-increasing technological innovations bring back the interest of Web 3.0. The core technologies that will be expected as building blocks of Web 3.0 may include blockchain for decentralization, IoT and 5G for communications, semantic extraction of web content, and extended reality for immersive human-computer experiences, as illustrated in Fig.~\ref{fig:web1toweb3}. The convergence of decentralization, semantic extraction, and extended reality toward Web 3.0 is closer to reality than ever, thanks to the development of emerging technologies such as AI, blockchain, Virtual reality (VR)/Augmented reality (AR), 5G, IoT, and edge computing~\cite{zarrin2021blockchain, issa2015artificial, adedugbe2020leveraging, drakatos2021triastore}. 

AI-driven algorithms can provide users with personalized and relevant information based on their behavior and preference, overcoming the information overload issue \cite{chen2022digital}. 
With recent advances in natural language processing, e.g., via large language models, AI-empowered agents, e.g., AI-based Decentralized autonomous organizations (DAOs), and AI-based VR interactive interfaces, are increasingly becoming a vital part of the future Web 3.0.
Additionally, the decentralization aspect of Web 3.0 aims to reduce reliance on central authorities, enhance security, and provide users with more control over their own data. For instance, blockchain technology can be implemented to ensure secure and transparent storage and sharing of personal data, giving users more ownership of their data \cite{ren2023building} \cite{tanwar2022blockchain}. It allows the users to control their information, countering centralized companies that currently dominate the majority of the web we use and interact with~\cite{zhao2020blockchain}. 

Web 3.0 provides many benefits and advancements over its predecessor, (i.e., Web 2.0, as illustrated in Fig.~\ref{fig:web1toweb3}). The key improvements of Web 3.0 are described as follows:

\begin{itemize}
    \item \textit{Semantic understanding and contextualization}: Web 3.0 introduces semantic technologies that enable computers to understand the context and meaning of data. Unlike Web 2.0 where semantic understanding is limited to the keywords and metadata associated with a type of content, semantic understanding in Web 3.0 is more sophisticated. The deeper understanding of data is achieved through AI and Machine learning (ML) algorithms, e.g., through large language models. This allows the web engine to make better outcomes and search results. The recent breakthroughs of large language models like OpenAI's ChatGPT and Google's Bard are the emerging applications of such semantic understanding based on the web's language, in which the search engine can understand the user's intent and return more relevant results.
    \item \textit{Decentralization and data ownership}: Unlike the centralization of Web 2.0 platforms, Web 3.0 emphasizes decentralization using blockchain technology such as Ethereum. This reduces the control of large corporations and gives users more ownership and control over their data. For example, NFTs can be utilized as unique digital assets to represent ownership of digital goods, e.g., digital arts, tickets, and memberships.
    \item \textit{AI-driven decision making}: With semantic understanding and AI integration, Web 3.0 supports innovative interactive applications based on VR/XR in which the user quality of experience can be optimized via learning algorithms, e.g., deep reinforcement learning \cite{zhang2023ai}. For example, an edge computing server running Web 3.0 services can adaptively control the average bitrate, rebuffering time, and inter-quality variation based on feedback from the users, resulting in higher resource utilization. The collaborative training between the edge servers, e.g., via federated learning, can further facilitate the secure and personalized information exchange between the users \cite{chen2020federated}. 
    \item \textit{Interconnected devices and IoT}: Web 3.0 extends beyond traditional web browsers to include IoT devices where the IoT devices are increasingly becoming a part of the Web 3.0 ecosystem. The IoT smart devices, such as smartwatches, VR/XR controllers, and smart home access points, provide data collection at a large scale in a decentralized manner. Furthermore, IoT devices can be used to automate tasks in Web 3.0 such as executing smart contracts or managing resource distribution.
    \item \textit{Data interoperability}: Semantic technologies of Web 3.0 can facilitate data interoperability, allowing heterogeneous systems and platforms to communicate and share information with each other. The interoperability can be achieved by various approaches, such as Thing Description, which will be described in detail in Section \ref{sec:iot}. Further incentive mechanisms with blockchain networks can engage the data owners to share and join the system, thus enhancing the benefits of the entire decentralized community.
    \item \textit{Privacy and security}: Decentralization and cryptographic techniques in Web 3.0 contribute to enhance privacy and security. Indeed, decentralization, e.g., blockchain, solves some traditional security problems but has also introduced new security threats, such as 50\% attacks~\cite{investopediaattacks}.  
\end{itemize}

\subsection{The Architecture of Web 3.0}

\begin{figure*}
    \centering
    \includegraphics[width=0.85\linewidth]{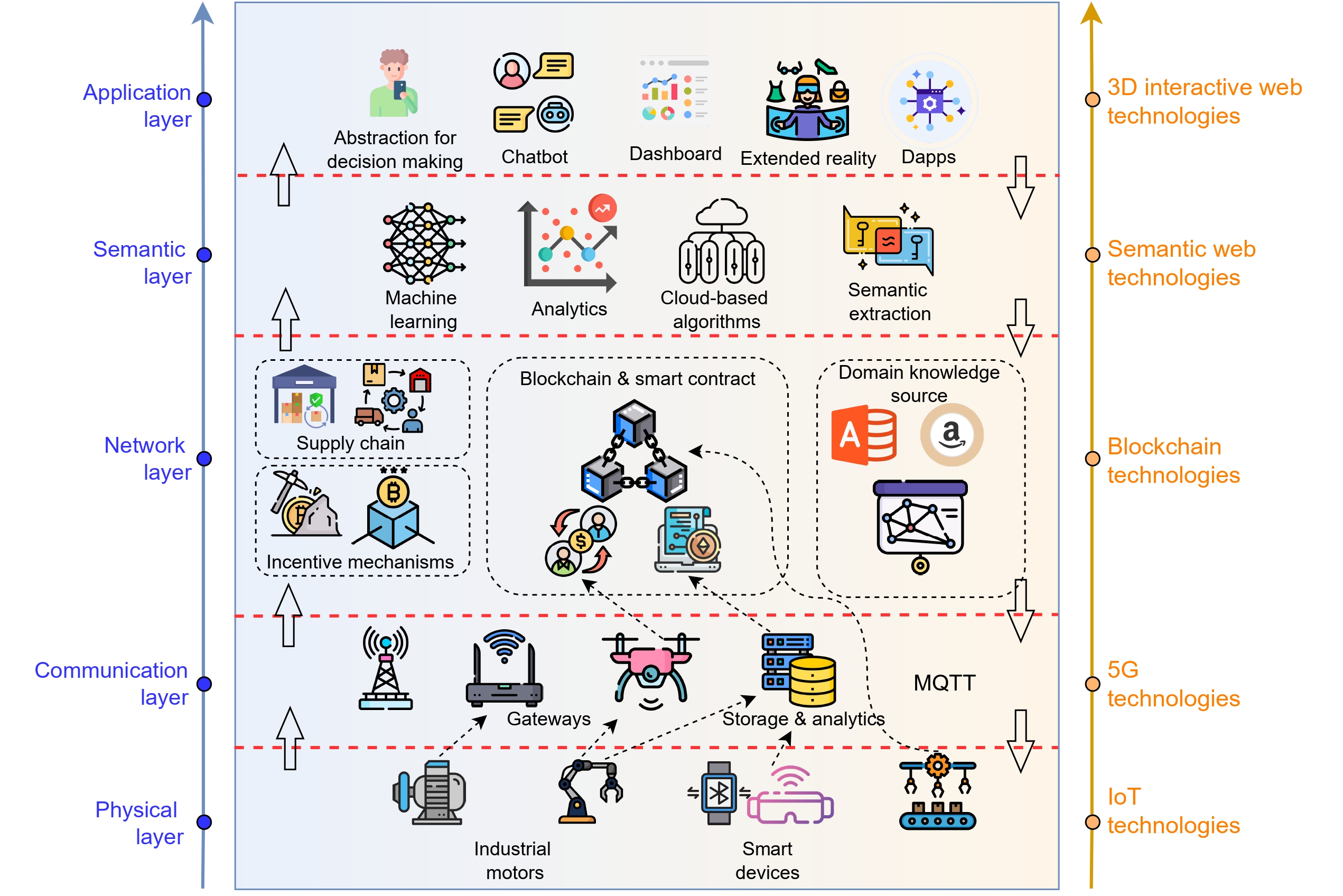}
    \caption{The architecture of Web 3.0 with five main layers. The data is communicated from the physical layer to the application layer through intermediate layers, i.e., the communication layer, network layer, and semantic layer. Each layer can be realized by corresponding technologies that are IoT, 5G, blockchain, semantic web, and 3D interactive web technologies.}
    \label{fig:web3_layout}
\end{figure*}

The architecture of Web 3.0 represents an evolution in the design and functionality of the Internet in which the Internet is founded on decentralization and powered by emerging technologies. In this work, we propose a comprehensive architecture for Web 3.0 with five layers as illustrated in Fig.~\ref{fig:web3_layout}.

\subsubsection{Physical Layer}

The physical layer is one of the important layers of Web 3.0 architecture which is responsible for data collection and interaction. A vast collection of smart devices and industrial motors are utilized at this layer to collect a wide range of data from the real environment. These devices allow the seamless integration of real-world data into the digital sphere, hence improving knowledge of different platforms and environments by efficiently collecting, processing, and transmitting data~\cite{drakatos2021triastore}. More details about IoT technologies used in Web 3.0 will be discussed in Section~\ref{sec:iot}.

\subsubsection{ Communication Layer}
 The communication layer connects multiple nodes using various protocols that allow decentralized and peer-to-peer data transmissions. It determines how data is packetized, addressed, transferred, routed, and retrieved~\cite{ruchika}. It includes critical components like network propagation and validation protocols, network security, privacy, and trust mechanisms. It efficiently shares data via propagation protocols and maintains data integrity through validation protocols. Network security measures defend against illegal access and assaults, while privacy controls provide users with the ownership of their personal information. These components provide a solid infrastructure that allows secure, efficient, and trustworthy data transfer in a Web 3.0 environment. To reliably deploy Web 3.0 in existing infrastructures like the 5G networks, several requirements of the communication layer need to match the 5G standards, such as network slicing, edge computing, and massive Multiple-Input Multiple-Output (MIMO). Further discussion about the 5G technologies for Web 3.0 will be presented in Section \ref{sec:5G}.

\subsubsection{ Network Layer}

The network layer is the most important layer of the Web 3.0 architecture which provides a decentralized, secure, and intelligent web. It improves data decentralization and trustless transactions while maintaining data integrity and security by integrating blockchain technology and smart contracts. Moreover, security is maintained using consensus mechanisms, ensuring that all nodes in the network agree on the state of the network eliminating the dependency on centralized authority~\cite{lashkari2021comprehensive} \cite{nguyen2019proof}. In addition, incentive mechanisms can be deployed at this layer to compensate users for their contributions, such as data storage or transaction validation~\cite{sangeet}. This layer aims to foster trust and transparency by reducing fraudulent activities. Furthermore, it builds a strong network platform resistant to censorship and single points of failure, resulting in a more open and trustworthy online environment. Blockchain technologies used in Web 3.0 will be discussed in more detail in Section~\ref{sec:blockchain}. 

\subsubsection{ Semantic Layer}

The semantic layer is responsible for analyzing and processing data. It is used to make predictions based on user activities that can be used to improve the user experience~\cite{salim2022data}. Several strategies and operations are available for the semantic layer. The operations can be cloud-based algorithms, machine learning, analytics, and semantic extraction. These approaches are all used to gather information from various sources, such as IoT devices.  For example, it can enhance users personalizing by providing recommendations based on user preferences, behavior, and interests. At this layer, semantic web technologies play the key role and they will be discussed in more detail in Section~\ref{sec:SWT}.

\subsubsection{Application Layer}

The application layer is the topmost layer of the Web 3.0 architecture which creates a user interface for decentralized applications (DApps) that run on top of blockchain networks. DApps are more secure and efficient than traditional centralized applications~\cite{liu2020graph}. It is built on open protocols and standards, allowing users to engage directly with services without the use of middlemen\cite{yue2021survey}. The application layer in Web 3.0 provides users with improved control over their data and digital identities. Through user-centric design and decentralized architecture, it enables personalized experiences, enhanced AI-driven services, and faster transactions, while simultaneously encouraging better privacy and security. At this layer, interactive 3D technologies play a key role in enhancing users' experiences and engagement in Web 3.0, and this will be discussed in more detail in Section~\ref{sec:3D_web}.

The architecture of Web 3.0 marks significant changes from conventional client-server paradigms. Its emphasis on decentralization, semantic data comprehension, open standards, user empowerment, and smart contract integration establishes the foundations for a more inclusive, interconnected, secured Internet environment~\cite{Georgia}. As Web 3.0 continues to develop, it has the potential to transform industries and open up new opportunities and applications for innovations and engagement. The next section explores the diverse applications of Web 3.0.

\subsection{Potential Applications and Services of Web 3.0}
{\color{black}

Although Web 3.0 is still in its initial stages, its technological foundation is rich. By leveraging cutting-edge technologies, Web 3.0 can offer numerous novel applications and services to build a more decentralized, secure, and user-centric Internet. This subsection discusses potential applications and services of Web 3.0.

\subsubsection{Decentralized Applications (DApps)}

DApps are important applications of Web 3.0. They exploit the potential of decentralization to change the way users engage with online resources. Unlike traditional apps that rely on centralized servers, DApps run on blockchain networks to distribute data and functionalities over a network of nodes~\cite{liu2020graph}. By doing so, it can mitigate the single-point-of-failure problem, enhance security, and give users more control over their data. Consequently, DApps provide an innovative level of openness, autonomy, and trust in the world of technology and cutting across various industries.

\begin{itemize}

\item \textit{Social media}: Web 3.0 social networking sites present a decentralized and user-centered substitute for conventional social platforms. Social media platforms have evolved into DApps that prioritize user ownership of data, privacy, and authenticity \cite{liu2020graph}.  It supports three-way communication and utilizes a decentralized network,  to give users more power and privacy. The primary purpose is to tackle the issues of privacy, data ownership, and censorship that centralized social media platforms frequently face~\cite{salim2021blockchain}. Steemit (www.steemit.com), Sapien (www.sapien.network), and Sola (www.sola.network) are a few notable examples of decentralized social media applications.

\item \textit{Decentralized gaming platforms}: Decentralized gaming platforms allow the players to engage with the games in innovative and decentralized ways. This represents an evolution beyond traditional gaming platforms to decentralized gaming with complete user ownership. It utilizes the blockchain to establish genuine ownership of in-game assets, which allows secure and transparent buying, selling, and trading processes~\cite{akash}. Through smart contracts, dynamic player-driven ecosystems can emerge, adapting game rules based on community agreement. Furthermore, Web 3.0 emphasize on interoperability supports cross-game experiences and Metaverses, allowing seamless movements between virtual worlds. Examples of decentralized Web 3.0 gaming applications include Decentraland (www.decentraland.org) and RaceFi (www.racefi.io).

\item \textit{Decentralized healthcare platforms}: Web 3.0's architecture brings groundbreaking advancements to healthcare systems. It revolutionizes the medical landscape by integrating advanced technologies and decentralized networks. This paradigm shift enhances patient-centered care through seamless data sharing, secure patient control over personal health records, and improved interoperability among healthcare providers\cite{chondrogiannis2022using}. Smart contracts ensure transparent and automated processes, while AI-driven diagnostics and telemedicine foster efficient, and personalized treatments~\cite{abhishekforbes}. AiGIA (www.aigia.health), and DeHealth (www.dehealth.world) are examples of decentralized Web 3.0 healthcare applications.

\item \textit{Decentralized education platforms}: Web 3.0 has the potential to revolutionize the education sector through its decentralized architecture and innovative technologies. It can make the e-learning platforms more individualized, accessible, and secure. Users have more control over their learning data thanks to decentralized systems, which can also guarantee privacy and ownership  \cite{li2019edurss}. Intelligent algorithms modify information to fit different learning preferences, increasing understanding and engagement. The reliability of certificates and credentials is ensured by blockchain technology, which reduces the risk of fake credentials\cite{siddiqui2022blockchain}. These platforms, run on a blockchain network, allow students/teachers to access information securely\cite{marjit2020towards}. The advantages include increased engagement, personalized learning paths, and globally recognized credentials, resulting in more flexible and globally aware learners ready to tackle future challenges. BitDegree (www.bitdegree.org) and TeachMePlease (www.teachmeplease.co.uk) are two examples of decentralized learning applications.

\end{itemize}

\subsubsection{Decentralized Finance (DeFi)}

DeFi is a revolutionary transition from existing financial systems, providing a variety of financial instruments such as borrowing, lending, and trading without the use of traditional middlemen~\cite{cao2022decentralized}. It employs blockchain technology to establish an open and permissionless financial services environment. Unlike existing systems that rely on middlemen, it is based on smart contracts, allowing for direct peer-to-peer transactions~\cite{ren2023building}. This reduces expenses, increases transparency, and promotes global financial access. DeFi users have more control over their assets, less reliance on middlemen, cheaper costs, and improved financial autonomy~\cite{ray2023web3}. Uniswap (www.uniswap.org), and Aave (www.aave.com),  are among many examples of applications of DeFi.

\subsubsection{Decentralized Autonomous Organizations (DAOs)}

DAO is a groundbreaking concept that represents a novel way of organizing and governing activities without relying on centralized authorities. In contrast to conventional hierarchical institutions, DAOs operate using blockchain-based protocols and smart contracts. Unlike centralized systems, it allows users to make collective choices and regulate the organization's operations without the need for middlemen which promotes openness, eliminates complexity, and increases inclusion~\cite{wang2019decentralized}. It improves efficiency, reduces corruption, and fosters global collaboration, improving governance standards~\cite{ding2022desci}. Some of DAO's applications are MolochDAO (www.molochdao.com), USDT (www.kucoin.com/trade/DAO-USDT), and MakerDAO (www.makerdao.com/en).

\subsubsection{Non-Fungible Tokens (NFTs)}

NFTs have emerged as a groundbreaking concept, reshaping the landscape of digital ownership and creativity. They mark a significant change from established ownership structures and unique digital assets or tokens. These unique tokens leverage blockchain technology to certify the authenticity and ownership of digital assets, such as art, collectibles, and virtual real estate~\cite{wang2021non}. Each NFT is different and indivisible thanks to blockchain technology~\cite{murray2023promise}. NFTs transform ownership in the digital era by offering monetization alternatives for creators and artists through smart contracts, royalties, and transparent provenance. Some well-known examples of NFTs are SuperRare (www.superrare.com), Nifty Gateway (www.niftygateway.com), and Async Art (www.async.art).

\subsubsection{Supply Chain Management (SCM)}

With the emergence of Web 3.0, this area has undergone significant transformations over the years. Unlike traditional systems often involving centralized control, decentralized supply chains provide transparent and traceable monitoring of items from source to consumer. It decreases fraud, ensures product authenticity, and improves overall efficiency. Decentralized tracking enables organizations and corporations to monitor goods and items from origin to destinations, granting greater control over their supply chain~\cite{benvcic2019dl}. This transformation revolutionizes inventory management, logistics, and traceability, leading to increased accuracy, decreased overhead costs, and enhanced trust among supply chain partners~\cite{kayikci2022blockchain}. Notable decentralized supply chain applications include VeChain (www.vechain.org), Provenance (www.provenance.io), and Tefood (www.te-food.com).

\subsubsection{Decentralized Identity Management}

Digital identity, especially the concept of self-sovereign identity, plays a vital role in the development of Web 3.0. It empowers users with ownership and control of personal information, without intermediaries~\cite{wang2023linking}. In conventional systems, user information was stored in a centralized database, however, this issue can be meditated using decentralized identity~\cite{avellaneda2019decentralized}. It makes user data more secure and reduces the chance of being stolen. It also removes the necessity for multiple usernames and passwords, enhancing the online experience and stopping identity theft~\cite{vcuvcko2021decentralized}. Moreover, it builds trust, simplifies proving a user's identity, and maintains privacy on various websites and apps. Dock DID (www.dock.io) and Onyx (www.jpmorgan.com/onyx/index) are two examples of decentralized identity management systems.

\subsubsection{The Metaverse}

This is a new platform transforming the connected digital world. It offers a fully immersive, interactive, and digital world that is connected via the Internet and accessible through VR and AR devices. The Metaverse enables seamless transitions between the physical and digital worlds, thus fostering creative expression and immersive interactions~\cite{hoang2023}. In the conventional virtual world concept, the user data is often stored in a centralized database, possibly leading to the point-of-failure problem. In this context, Web 3.0 technology can transform the Metaverse into a decentralized network, e.g., Decentraland (www.decentraland.org). In this way, Web 3.0 can play a pivotal role in enabling the metaverse using blockchain, semantic, and interactive 3D web technologies. Thus, it ensures that digital assets, identities, and transactions are secure, verifiable, and owned by the users themselves~\cite{liu2023slicing4meta}. In the Metaverse, virtual assets are truly owned by users, who can freely trade and exchange them. Blockchain technology secures this ownership, making it unique and tradeable without middlemen. Decentraland and Somnium Space (www.somniumspace.com) are a few well-known examples of Web 3.0-enabled Metaverse systems.

\subsection{Current Standards and Industry Development}

Many international companies and organizations are actively developing standards toward Web 3.0.

\subsubsection{The World Wide Web Consortium (W3C)}
The W3C (www.w3.org) is an international organization that has been developing open standards for the WWW, such as HTML, CSS, and XML, and it continues to play an important role in Web 3.0.  By establishing guidelines for semantic data representation, linked data principles, and interoperability it guarantees a unified and organized framework for Web 3.0 development (www.w3.org/standards). These standards support machine-readable data, allowing intelligent agents and systems to effectively grasp and process information. W3C aims to build a seamless and linked digital world in which data can be easily exchanged, recognized, and used across multiple platforms, improving the capabilities and possibilities of the next-generation Internet.

\subsubsection{The ConsenSys}
ConsenSys (www.consensys.io) stands as the leading blockchain and Web 3.0 software companies, driving the forefront of innovation in decentralized technologies and reshaping the digital landscape. It plays an important role in defining and promoting Web 3.0 principles and blockchain-based solutions. ConsenSys takes part in several other initiatives that are pushing the development of Web 3.0. For example, it focused on its business of building DeFi apps, which are financial solutions based on blockchain technology. ConsenSys is also interested in developing NFTs, unique digital assets that cannot be reproduced~\cite{consensys}.

\subsubsection{The InterPlanetary File System (IPFS)}

IPFS (www.ipfs.tech) stands as a trustworthy industry standard within the Web 3.0 landscape. It transforms existing web infrastructure by operating decentralization and distributed content sharing. It uses a content-addressable method to get material based on its encrypted hash rather than depending on centralized servers with specific uniform resource locator (URLs)~\cite{murray2023promise}. As such, IPFS improves data availability and fault tolerance while lowering the danger of single points of failure. IPFS promotes a more open and robust online by enabling users to exchange and access material without the necessity of intermediaries. In addition, IPFS measures with the idea of a decentralized into several data ownership, security, and accessibility are key, leading to a new age of several online interactions~\cite{ipfs}.

\subsubsection{The Solid}

The idea of Solid (www.solidproject.org), developed by Sir Tim Berners-Lee, the inventor of the WWW, plays an essential role in developing Web 3.0. It addresses concerns about data ownership, privacy, and control by providing ``pods'' personal data repositories that provide users with data authority and transparent sharing across apps. This method solves existing privacy and centralization challenges. The relevance of Solid derives from its coherence with the decentralized concept of Web 3.0 and its ability to transform the digital environment. It incorporates the ideas of a more user-centric, secure, and open internet by stressing user control and cross-platform data exchange, proposing itself as a feasible standard for the Web 3.0 paradigm.

\subsubsection{The Polkadot}
Polkadot (www.polkadot.network) is a decentralized Web 3.0 concept that aims to realize the decentralized web's multichain concept. It separates the rest of the third-generation advanced blockchains with characteristics like real interoperability, para chains, para threads, great energy efficiency, and user-driven governance. This ecosystem enables the creation of unique DApps and solutions that can serve a wide range of Web 3.0 applications.

\subsubsection{Industry Development}
Web 3.0 is bringing in a new age of industrial growth, marked by innovative technologies and decentralized frameworks that are transforming traditional industries. This transformative revolution is through the integration of blockchain, smart contracts, and DApps, which enhance security, and efficiency in various industries. The following are the most recent Web 3.0 industry developments:

\begin{itemize}

\item \textit{Everledger}: Everledger (www.everledger.io) is a Web 3.0 concept that provides technical solutions to promote openness in global supply chains. It supports tracing an item's use throughout the entire cycle of the global supply chain. For authenticity verification, customers can scan the label. The Everledger software establishes a significant milestone in web security by offering fraud protection without the need for sensitive personal information or challenging passwords. 

\item \textit{The Storj}: Storj (www.storj.io) is a decentralized storage system that allows users to store information securely while also assuring both redundancy and fault tolerance. It uses blockchain technology to transform the benefits of cloud storage, considering distributed networks. The most notable advantage of Storj is the ability to do limitless uploads and downloads at any time. Another notable aspect of Storj as a Web 3.0 application is the Storj currency, which supports the Storj decentralized storage infrastructure.

\item \textit{The Uniswap}: Uniswap (www.uniswap.org) is a Web 3.0 exchange mechanism that employs an open and decentralized network protocol to provide users ownership. It enables users to exchange any Ethereum request for comment (ERC-20) token with no middlemen, fees, or verification of identity. It also allows users to supply stability to the marketplace and receive fees in exchange.

\item \textit{The Axie Infinity}: Axie (www.axieinfinity.com) Infinity is an innovative Web 3.0 gaming platform that encourages players to participate in games, earn rewards, and exchange NFT-based game assets. To utilize these services, users must go through a multi-step procedure that includes establishing an account and syncing their wallet for quick access.

\end{itemize}


\section{IoT Technologies for Web 3.0}
\label{sec:iot}
\subsection{Background}
\label{subsec:iot-background}

The IoT has experienced exponential growth due to advancements in cutting-edge technologies, such as 5G, edge computing, AI, and blockchain. According to current projections, approximately 34.7 billion IoT devices will be commercially launched by 2028, with an anticipated annual growth rate of around 18\%~\cite{ericsson}. These interconnected and heterogeneous devices autonomously generate and exchange vast amounts of data without manual intervention. The integration of IoT technology into Web 3.0 is expected to revolutionize the digital environment, enhancing its connectivity and intelligence. By incorporating real-life data from various types of IoT devices, including smart appliances, smart vehicles, and wearable sensors, Web 3.0 has the potential to provide users with a more human-centric and seamless experience. The ubiquity of IoT devices also presents opportunities for innovative applications, such as digital twins and the Metaverse, thereby fostering future cyber-physical systems~\cite{wang2022metaverses}.

\begin{figure}[h]
    \centering
    \includegraphics[width=0.99\linewidth]{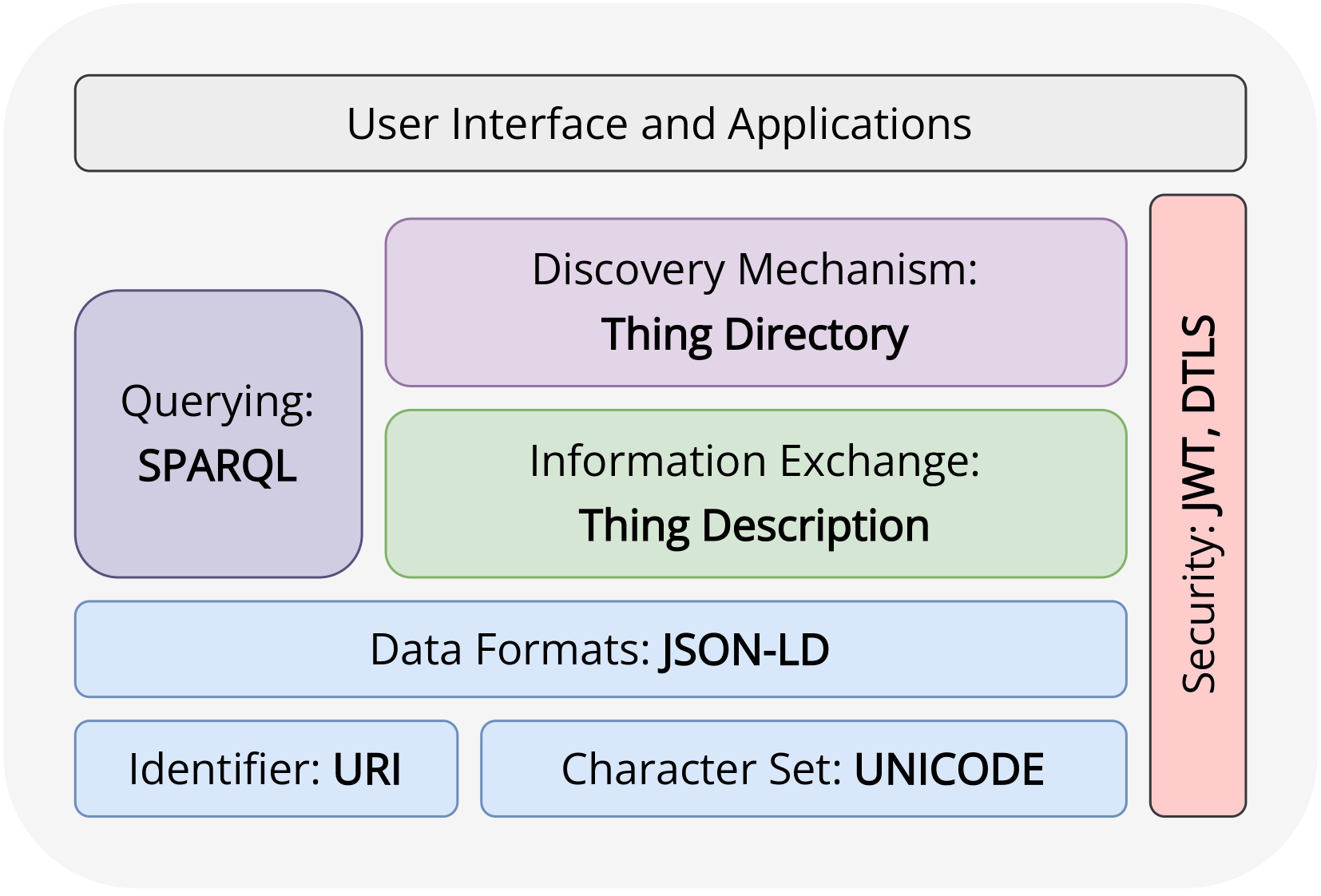}
    \caption{An overview of the W3C's Web of Things (WoT) architecture~\cite{novo2020semantic}. The architecture hinges on key components: \textit{Things}, encompassing physical and virtual entities, detailed by \textit{Thing Descriptions} that standardize capabilities and interactions. The \textit{Thing Directory} facilitates efficient discovery and access to Things through URIs. SPARQL enhances querying capabilities, while the JSON-LD data format ensures interoperability and semantic understanding.}
    \label{fig.WoT_stack}
\end{figure}

However, the continuous development of the IoT ecosystem has led to increased complexity in data compatibility and system interoperability. This complexity arises from the wide variety of IoT devices, sensing data, and deployment contexts, which are categorized by different manufacturers and service providers. Additionally, the scale in both volume and heterogeneity across various application fields continues to grow, further limiting the autonomous exchange of data and service interactions between different IoT systems. Addressing these challenges requires the establishment of standardized protocols and frameworks that enable seamless communication and collaboration between diverse IoT devices and service providers. Such standardization processes would facilitate the integration of new devices and applications, improve data security, and enhance the overall performance of IoT systems, making them more reliable and efficient.

\subsection{Established Frameworks and Standardization}

Several standardization organizations have developed new protocols and platforms to integrate the semantic web and IoT technology. For example, a Semantic Web of Things for Industry 4.0 (SWeTI) architecture is proposed in~\cite{patel2018raw}. The SWeTI architecture with several layers to enable interoperability across IoT devices and services. In this way, the semantic information can be extracted from lower layers, e.g., the physical layer, and utilized by upper layers, e.g., the semantic layer and application layer.
Other approaches have been proposed to standardize architecture for Web 3.0-based IoT technologies. The \textit{Web of Things} (WoT) is a concept established by the W3C to enable interoperability across IoT devices and systems~\cite{w3c_wot}. The WoT utilizes and extends existing web technologies for simplifying the development of IoT services and applications. In general, WoT presents a basic interaction abstraction that relies on IoT devices' properties, events, and actions. This abstraction provides a universal reference point for applications to obtain metadata for an IoT service and to comprehend how to access the data and functions of the IoT service. The WoT also enables linked data-based semantic annotations that provide robust search and inference capabilities. Therefore, the WoT architecture provides a platform for establishing a more integrated and interoperable IoT ecosystem.

\subsubsection*{Semantic Technologies in the WoT}
\label{subsec:iot-wot}
The WoT framework utilizes the semantic web technology standards to enable communication capabilities for IoT devices. An overview of the W3C's WoT architecture is illustrated in Fig.~\ref{fig.WoT_stack}~\cite{novo2020semantic}. In this architecture, \textit{Thing Description} (TD) is a key component that enables interoperability and communication between IoT devices and applications. The TD is not tied to any particular configuration, communication protocol, or application domain. TDs provide a standardized way of describing the metadata and interfaces of Things (i.e., IoT devices) in a machine-readable format, making it easier for other smart devices and applications to discover and interact with them. TDs typically contain information about the device's properties, actions, events, and interactions, as well as the data schemas, communication protocols, and formats. In particular, properties specify the state of Things, e.g., temperature or luminance value of a sensor. Actions define the functions that IoT devices can perform to manipulate their states such as changing a control heating/cooling strategy or toggling a light switch. Events describe event conditions that asynchronously push notifications and event data to a group of subscribed clients, e.g., overheating warnings or low level of lighting. 

\begin{figure}[h]
    \centering
    \includegraphics[width=\linewidth]{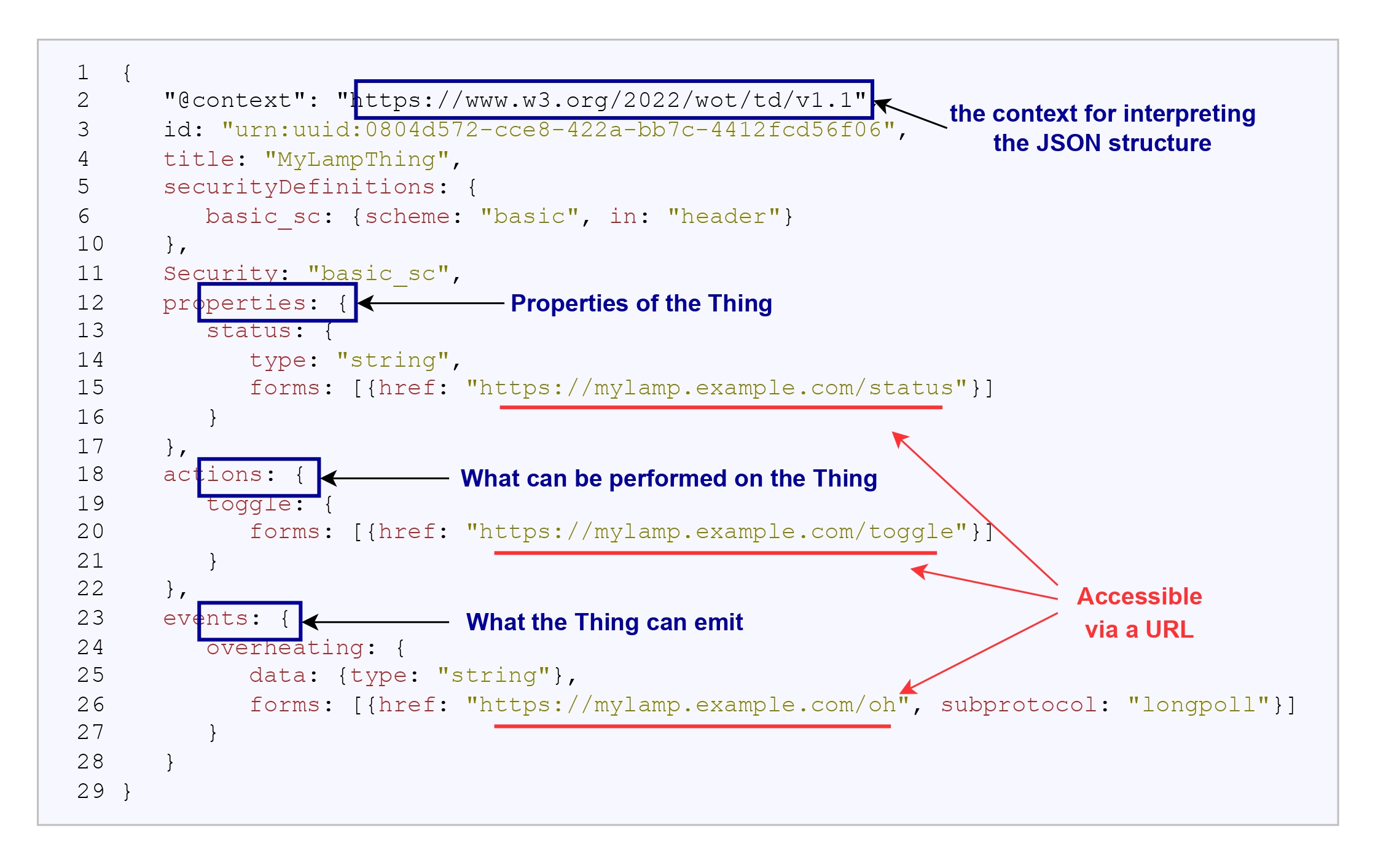}
\caption{A sample JSON structure of a TD~\cite{hiteshjethva}.}
\label{fig.TD_JSON}
\end{figure}

The underlying data format for the TD is Resource description framework (RDF), which is the basic metadata format for data exchange in the semantic web. By default, TDs are encoded in JavaScript object notation (JSON) format. Fig.~\ref{fig.TD_JSON} illustrates a sample JSON structure of a TD. It includes the Thing's name, a unique identifier, a human-readable title, security configurations (i.e., for authentication, authorization, etc.), all the interactions that the Thing can support, and an optional description. To allow extensions and extensive semantic analysis, the JSON serialization of TDs also supports the syntax of the JSON-LD format for Linked Data. The use of JSON-LD format can facilitate semantic processing by converting a TD into RDF triples, performing semantic inference, and executing tasks based on ontological terms. This enhanced processing ability can promote greater autonomy among clients.

A repository for TD instances, called \textit{Thing Directory}, allows for registering, updating, and removing TDs, as well as performing lookup operations through SPARQL queries. When a device retrieves the TD of another IoT device from a Thing Directory, it must interpret and comprehend the information contained in the TD to establish a successful interaction. Once this is achieved, the devices can communicate with each other autonomously without requiring any human intervention.

\subsection{The Roles of IoT Technologies in the Future Development of Web 3.0}

The IoT is a widely recognized concept referring to a network of infrastructures that connect embedded devices and objects, such as sensors, vehicles, and appliances, via the Internet. This connectivity enables the exchange and processing of data. When combined with Web 3.0, IoT technology has the potential to significantly transform our interactions with devices, data management and utilization, and overall digital experience. These transformations have resulted in more adaptable processes within an open environment that promotes collaboration, shared services, and group decision-making.

\subsubsection{Decentralized Data Ecosystem}

Web 3.0 is built on the principle of creating a decentralized data ecosystem. This vision aims to empower users by granting them more control over their data and enabling direct interactions with applications and services. To achieve this, decentralized solutions play a crucial role in providing scalability benefits and addressing the vulnerabilities associated with relying on single points of failure. IoT devices are key contributors to this vision as they generate vast amounts of real-time data at the network's edge. By integrating IoT-generated data into blockchain networks, the ownership and control of data are distributed among users, leading to a more transparent and reliable digital environment.

\begin{figure*}[!]
    \centering
    \includegraphics[width=0.90\linewidth]{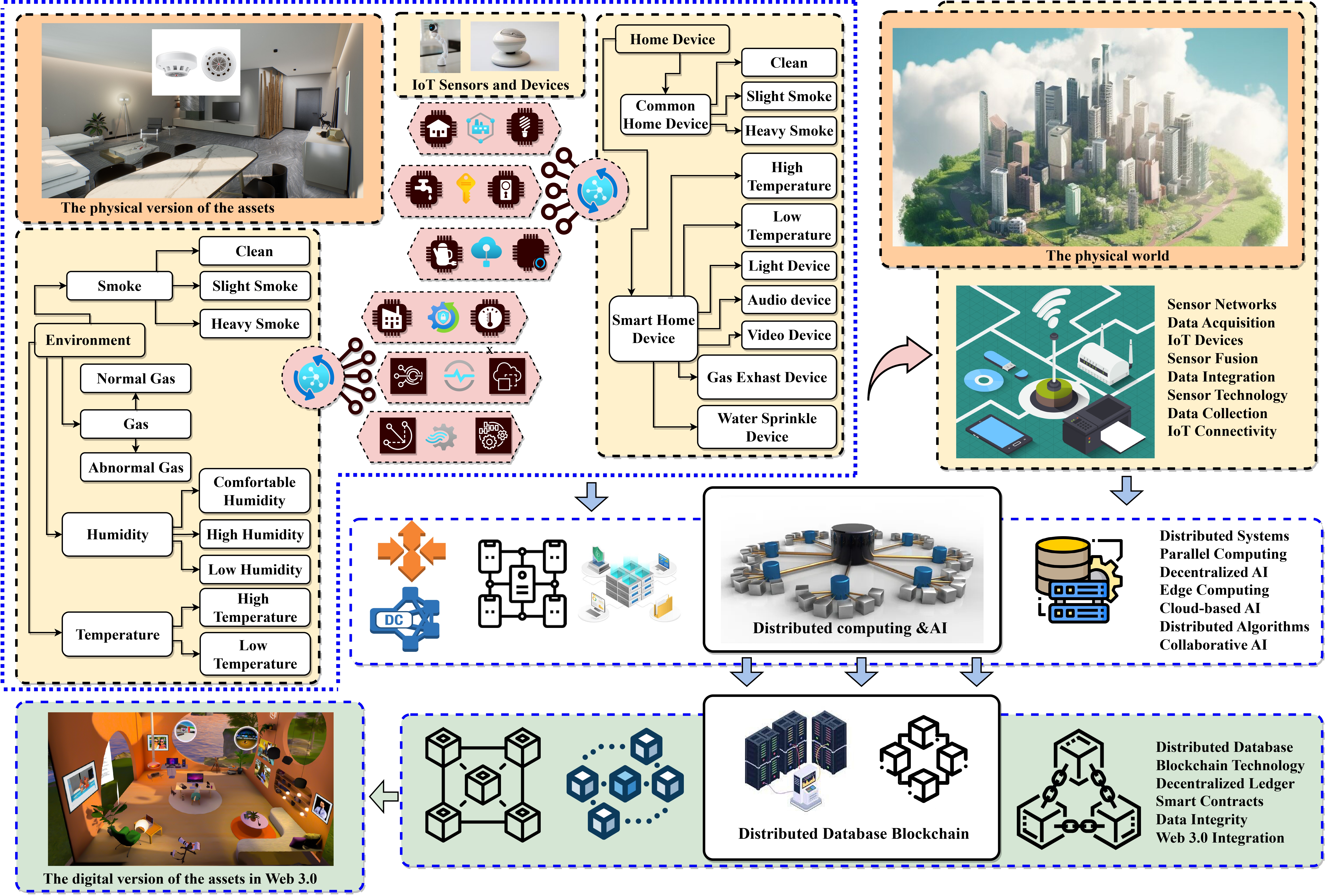}
\caption{Illustration of a cutting-edge multi-layer cloud architecture in Web 3.0, showcasing IoT integration. Smart Home concept exemplifies the rising importance in highly developed areas, overcoming challenges with distributed computing and blockchain \cite{tao2018multi}.}
\label{fig.IoTTech}
\end{figure*}

\subsubsection{Enhanced Security and Privacy}

Security and privacy are critical in Web 3.0, and IoT offers advantages in these areas. IoT devices can use end-to-end encryption, anonymization, secure data transmission, and blockchain's tamper-proof ledger to protect sensitive data from unauthorized access and manipulation \cite{rahman2020secure,saxena2021blockchain, farahani2021convergence}. This increased level of security gives users more assurance and faith when engaging with the digital realm, even in scenarios where individuals might consider trading their private data for rewards. 

\subsubsection{Empowering User-Centric Interactions}

IoT plays a crucial role in fostering user-centric interactions within Web 3.0. It enables personalized, context-aware, and autonomous experiences. A human-centric IoT design focuses on users' requirements in terms of functionalities and interaction modalities \cite{koreshoff2013approaching}. This shift towards user-initiated adoption of near-edge fog architectures for computation and storage \cite{oteafy2018iot}. Through IoT-enabled smart devices, users can seamlessly interact with their digital environment. Whether it is a smart home adjusting temperature based on occupancy or a wearable device tracking fitness data, IoT empowers users to engage with technology naturally and intelligently \cite{wu2014cognitive}. This trend extends to personalized IoT systems for smart cities under fog computing architecture. This leads to improved service latency, reduced functional mismatches, better resource visibility, and lower energy consumption \cite{yannuzzi2014key}.

\subsubsection{Autonomous and Trustless Transactions}

When IoT devices are combined with blockchain-based smart contracts, they facilitate autonomous and trustless transactions \cite{wang2019blockchain} \cite{fernandez2018review}. These self-executing agreements eliminate the need for intermediaries, simplifying procedures and lowering transaction costs. This convergence improves efficiency, scalability, diversity, and reliability within the Web 3.0 framework, introducing new possibilities for smooth and secure digital interactions across various applications. These applications range from supply chain management, where real-time tracking and transparency are critical, to financial services, where transactions require enhanced security and trust. This critical combination represents a significant advancement in the evolution of the decentralized Internet, resulting in an upcoming age of unforeseen possibilities for innovation and value creation.

\subsection{Recent Research for Integrating IoT Technologies in Web 3.0}

Current research works in integrating IoT technologies and Web 3.0 mostly focus on semantic technology, blockchain, and interoperability. In this subsection, we highlight the most significant research findings in these domains, giving us a general direction of the latest advancements and their implications for the potential interconnected devices and the evolution of the Web 3.0.

The conceptual arrangement depicted in Fig.~\ref{fig.IoTTech} illustrates the integration of IoT technologies to enhance the immersive experience of Web 3.0. Using the example of the Smart Home concept, the growing significance of this idea in highly developed areas is attributed to the widespread network coverage and embedded computing technologies. Despite this, existing solutions exhibit heterogeneity due to various factors leading to the involvement of diverse services and devices. To address these challenges and leverage the opportunities presented by advancements in IoT and cloud computing, a new multi-layer cloud architectural model for Web 3.0 is proposed. In this model, the next layer involves the analysis of outputs from IoT technologies through distributed computing and AI. Subsequently, these outputs are transferred to a distributed database blockchain. Finally, the digital representation of assets in Web 3.0 is presented within a virtual environment.

\subsubsection{Enhancing interoperability amid IoT technologies in Web 3.0}

A recent study on interoperability concepts, focusing on IoT technologies and the emergence of Web 3.0, was presented in \cite{hatzivasilis2018interoperability}.
This work focused on addressing challenges related to achieving interoperability in the modern IoT ecosystem, identifying organizational, semantic, syntactic, and technological interoperability as key obstacles. The proposed solution introduced an interoperability framework, facilitating cooperation from device to backend, enabling seamless interaction across IoT layers. Categorized into four levels, the framework ensures technological, syntactic, semantic, and organizational interoperability, promoting seamless communication and collaboration. The presented practical roadmap can overcome integration challenges, aligning with standards and shared models. Implemented by the EU-funded SEMIoTICS project, it stands as a robust response to IoT and Web 3.0 interoperability issues. In addition, addressing interoperability challenges is crucial for the efficient operation of a Web 3.0 platform handling multisensory data from diverse IoT devices. These challenges, stemming from technical and organizational factors, encompass standardization, semantic technologies, data integration, and collaboration among stakeholders. A potential solution involves utilizing standardized protocols and semantic web technologies, ensuring consistent communication and meaningful data interpretation. Adopting Linked Data principles and ontologies further enhances data sharing~\cite{linkeddata2}. Industry standards like OPC-UA and PPMP provide platform-independent protocols, promoting real-time communication and uniform representation. Implementing these protocols, along with technologies like WoT's Thing Description, can enhance interoperability in a Web 3.0 environment~\cite{opc2008,ioteclipse,w3c_wot}.

\subsubsection{ Blockchain enabled IoT and edge technologies in Web 3.0}
In the context of IoT integration into Web 3.0, challenges with massive IoT workloads were addressed by proposing Triastore, an approved blockchain framework with ML deployed the edge devices~\cite{drakatos2021triastore}. Triastore abstracted ML models into approved information blocks, incorporating Blockchain Consensus and Proof of Federated Learning (PoFL) phases. The study's practical assessment using MNIST data demonstrated high precision with minimal data integrity degradation. Additionally, the paper introduced a collaborative consensus mechanism, PoFL, enhancing user security and privacy by sharing parameters among distributed parties. The proposed approach was seamlessly implemented with the Fabric open-source platform, enhancing the authenticity of blockchain assessments~\cite{drakatos2021triastore}. Semantic technologies contribute to addressing trust, security, and privacy challenges in IoT, enabling fine-grained control over data access and sharing by representing security policies, access controls, and privacy preferences. Semantic reasoning aids in enforcing security, anomaly detection, and ensuring data integrity and confidentiality in IoT deployments. However, applying semantic technologies to large-scale IoT deployments for Web 3.0 can pose scalability challenges, requiring significant computational resources and potentially introducing latency issues. Managing ontology consistency in dynamic IoT environments is difficult, but potential solutions include ontology design patterns, alignment and mapping, and ontology evolution management frameworks. Overcoming these challenges necessitates collaborative efforts, standardization, optimization techniques, and advancements in data quality. Interdisciplinary collaborations with AI and distributed systems can drive the development of robust and efficient Web 3.0 technologies, considering the integration of blockchain and IoT technologies~\cite{choi2019ontology, mozzaquatro2018ontology, tao2018multi, gheisari2021obpp, szilagyi2016ontologies}.

\begin{table*}
	\caption{Summary of Recent Research for Integrating IoT Technologies in Web 3.0}
	\label{table_IoTweb}
	\begin{centering}
		\begin{tabular}{|>{\raggedright\arraybackslash}m{0.5cm}|>{\raggedright\arraybackslash}m{1.8cm}|>{\raggedright\arraybackslash}m{3.0cm}|> {\raggedright\arraybackslash}m{4.3cm}|>{\raggedright\arraybackslash}m{6.0cm}|}
			\hline 
		\multicolumn{1}{|>{\centering\arraybackslash}m{0.5cm}|}{\multirow{1}{*}{\textbf{Ref.}}} & 
            \multicolumn{1}{>{\centering\arraybackslash}m{1.5cm}|}{\multirow{1}{*}{\textbf{Application}}} &
            \multicolumn{1}{>{\centering\arraybackslash}m{3.3cm}|}{\multirow{1}{*}{\textbf{Research problem}}} &
            \multicolumn{1}{>{\centering\arraybackslash}m{4.3cm}|}{\multirow{1}{*}{\textbf{Solution}}} &
            \multicolumn{1}{>{\centering\arraybackslash}m{6.0cm}|}{\multirow{1}{*}{\textbf{Result}}}\\
			\hline 
			\hline 
   
        \cite{hatzivasilis2018interoperability} &  General IoT application & Semantic IoT integration challenges: scalability and consistency issues &Semantic technologies and collaboration for IoT integration challenges & SEMiOTICS adoption enhances IoT and Web 3.0, fostering interoperability despite diverse vendor product usage.\\	
         \cline{1-5} 

        \cite{drakatos2021triastore} & Smart city & Addressing massive data workloads issues on Web 3.0 & Blockchain data storage using federated learning  & Significantly increased data quality, performance, and data storage security resulting in high accuracy (95\%) and minimal learning loss (10\%).  \\ 	
        \cline{1-5}

       \cite{ choi2019ontology} & IoT-cloud security service for power systems & Detect and prevent security vulnerabilities for power systems & Develop an intelligent security framework for power IoT-Cloud systems, utilizing ontology reasoning, semantic-web technologies, and advanced intrusion detection methods & The study proposes a security framework, including ontology, inference rules, and attack detection for power IoT-Cloud systems, demonstrating effective attack detection \\ 	
        \cline{1-5}

       \cite{ mozzaquatro2018ontology} & General IoT applications & IoT exposes sensitive data to cyber threats, demanding enhanced cybersecurity & Propose an ontology-based cybersecurity framework using knowledge reasoning for IoT, composed of two approaches, (i.e., dynamic method to build security services and real-time monitoring of the IoT environment) &  Introduced ontology-based IoT cybersecurity framework, validated with an industrial scenario, revealing strengths, challenges, and valuable contributions for improvement\\ 	
        \cline{1-5}

       \cite{ tao2018multi} & Smart home & Heterogeneity in devices, services, and standards hinders IoT-based smart home application & Propose a multi-layer cloud architectural model is developed for IoT-based smart homes, which provides a substantially improved degree of interactions/interoperations between heterogeneous home devices and services & Average response times without loads range from 23.68 ms to 52.36 ms. Security framework effectively prevents various known attacks\\ 	
        \cline{1-5}
   
       \cite{ gheisari2021obpp} & Smart city & Heterogeneity and privacy-preserving for IoT data &  Propose three-module framework to address heterogeneity issue while keeping the privacy information of IoT devices & OBPP resists data attacks in 2.5s, computational cost <10\%, complexity O(n) with a 4x coefficient, applicable to smart cities\\ 	
        \cline{1-5}
   
       \cite{ szilagyi2016ontologies} & General IoT applications & Growing number of IoT devices and their diversity & Review  some of the Semantic Web technologies used in IoT systems, as well as some of the well-accepted ontologies used to develop applications and services for the IoT & Integration of semantic technologies (i.e., interpretation layer, processing layer and services/applications layer) with adapted machine learning algorithms will produce a smarter IoT\\ 	
        \cline{1-5}
			\hline                                                       
	  	\end{tabular}
		\par\end{centering}
\end{table*}

 \textit{Summary}: Table~\ref{table_IoTweb} summarized some emerging research works discussed in this subsection. In summary, recent research in integrating IoT technologies into Web 3.0 mainly focuses on overcoming challenges through semantic information extraction and technologies. Collaborative efforts, standardization, and interdisciplinary collaborations are crucial for advancing robust and efficient technologies, e.g., blockchain, semantic web, 3D interactive, 5G, and IoT.

\section{5G Technologies for Web 3.0}
\label{sec:5G}

\subsection{Background of 5G Technologies}

5G is the fifth generation of wireless networks and it is believed to become a monumental leap in the area of connectivity. Recent statistics illustrate that by the end of 2022, nearly 35\% of the global population had accessed 5G and this figure is projected to reach 85\% by 2028~\cite{ericsson_5G}. Compared with the conventional network generations, 5G can deliver faster, more reliable and more efficient mobile communication. It is expected to provide new applications and services with high data rates, low latency, massive connectivity, enhanced bandwidth and seamless interconnectivity~\cite{xing20215G}. Typically, 5G technology can be defined by three exemplary use cases~\cite{imt2015vis}:
\begin{itemize}
    \item \textit{eMBB}: Enhanced Mobile Broadband is the primary approach to mobile communication. It aims to provide communication with high data rates and large traffic volumes. By utilizing eMBB, users can enable Web 3.0 services including gaming, 3D videos, 4K streaming, and VR and AR~\cite{xing20215G}. As a result, this technology can enhance user experiences in web environments significantly and promote the development of new interactive 3D web applications. 
    
    \item \textit{URLLC}: URLLC represents an approach in 5G communication which provides high reliability and low latency services, thereby enhancing real-time data exchange and processing in technologies such as IoT, IoV and smart city. Following 3GPP, URLLC is required to transmit a packet with 99.999\% for 32 bytes with a user plane latency of 1 ms~\cite{3GPP20225GS}. The implementation of URLLC in Web 3.0 enables seamless and instantaneous communication, improving the real-time performance of Web 3.0 applications such as decentralized finance, industrial automation, autonomous vehicles, etc.
    
    \item \textit{mMTC}: Massive Machine-Type Communication is another 5G use case that simultaneously supports the connection of many devices. The requirements of these devices are low complexity and extended power consumption for data transmission~\cite{imt2015vis}. Applying mMTC to Web 3.0 paves the way for a more interconnected and automated web. It mainly supports the physical layer of Web 3.0 which contains a vast number of IoT devices, enhancing the functionality of applications e.g., smart city, environmental monitoring, supply chain management, etc. 
    
\end{itemize}

By utilizing the advantages of 5G, Web 3.0 can transmit enormous amounts of data, thereby improving the user experience of existing applications and services. In practice, 5G has had a major effect on human lives. For instance, industrial IoT is the most popular technology that benefits from 5G. In this scenario, 5G can be used to connect multiple sensors and machines to the Internet, which can help businesses expand efficiency and correctness~\cite{Chet20205g}. Otherwise, the attributes of 5G can improve the upcoming applications such as ensuring safety by utilizing low latency for real-time deployment of autonomous vehicles and traffic management in smart cities~\cite{hakak2023av}. However, 5G is not a single technology. It combines various technologies and methods to create an advanced wireless communication system. Below are some of the 5G key enabling technologies:  
\begin{itemize}
    \item \textit{Network Slicing}: Network slicing is another pivotal technology of 5G, which creates multiple virtual networks on a single physical infrastructure, catering to different types of services and users. Even though network slices share the same physical environment, they operate independently, ensuring that the disruption or issues from one slice cannot interrupt others. Besides, resource management in network slices depends on its service. This approach not only maximizes 5G core networks but also diversifies different quality of services, thereby making modern wireless networks more adaptable and user-centric~\cite{3GPP20225GS}. In the Web 3.0 ecosystem, network slicing provides tailored network segments for diverse applications such as Metaverse and DeFi, ensuring optimal performance and security for the decentralized web environment.

    \item \textit{Edge Computing}: Edge computing is a key technology of 5G that enables data to be processed at the edge of networks instead of a centralized database. This approach can minimize the data stream in wireless traffic, thereby expanding the bandwidth usage and ensuring low latency. Moreover, edge computing provides better security and privacy. As data is closer to its source, the transmission of sensitive content is minimized~\cite{abbas2018mec}. This leads to future wireless technologies being more decentralized, faster, safer, and more responsive. By rendering and processing data near the edge, edge computing can boost the capabilities of augmented and virtual reality, delivering immersive and seamless experiences in Web 3.0.    
    \item \textit{Massive MIMO}: Massive MIMO plays a significant role in 5G, as it utilizes an extensive array of antennas at both the transmitter and receiver to enhance the capacity and performance of wireless networks without requiring more spectrum. Following that, the enormous number of antennas allows spatial multiplexing where base stations can broadcast multiple data streams simultaneously to single or multiple recipients with the same frequency channels~\cite{xing20215G}. As a result, by utilizing the potential of Massive MIMO, advanced technologies can be deployed in real-time with high spectral efficiency and throughput~\cite{xing20215G}. This technology is highly effective in enhancing mobile augmented reality with a high data rate network, thereby providing immersive content and real-time interaction in Web 3.0~\cite{siri2021survey}.
    
    \item \textit{Beamforming}: Beamforming is another essential 5G technique that can effectively control the energy and direction of wireless signals. Instead of broadcasting signals uniformly in all directions, the beamforming technique can manipulate the amplitude and phase of the signal from each antenna, thereby directing the energy of radio waves toward a specific direction or user. This not only enhances the quality and range of the radio wave but also optimizes the spectrum usage to adapt to the soaring number of wireless data~\cite{ah2018beam}. Due to the ability to manage signal direction and strength, beamforming can ensure reliable connections for the vast array of IoT devices, thus enabling instantaneous data collection and exchange for Web 3.0.

    \item \textit{Millimeter Wave}: Millimeter wave (mmWave) refers to a specific part of the electromagnetic spectrum with wavelengths ranging from 1 to 10 millimetres, corresponding to frequencies between 30 and 300 GHz~\cite{mezzavilla2018end}. Such high radio frequencies allow 5G to broadcast data over wide bandwidths and small wavelengths, resulting in a faster cyber environment~\cite{rap2015mm}. Utilizing the high-frequency capabilities of mmWave in wireless technologies can improve the communication speed and throughput in Web 3.0, addressing the demand for high bandwidth and massive data transmission. 
\end{itemize}

\subsection{Roles of 5G Technologies in Web 3.0}

5G technologies provide unprecedented speed, low latency, and large simultaneous connection between devices~\cite{xing20215G}. Meanwhile, Web 3.0 requires an advanced network connection to support extensive communication among technologies like Blockchain, IoT, and the Metaverse. Thus, 5G networks, which have already been crucial for the growth of IoT, AI, and extended reality, can significantly contribute to Web 3.0 development~\cite {zeeve2023web3}. The fusion of Web 3.0 and 5G has the potential to gain a more distributed, reliable, efficient and online web experience. Following that, users can participate in a web environment with instant data transmission, enhanced virtual interaction, and interconnected digital communication~\cite{ray2023web3}. The contributions of 5G technologies in Web 3.0 are outlined in the following subsections.

\subsubsection{Enhancing decentralized communication}
The main feature of Web 3.0 is the decentralized web environment. By distributedly managing data, Web 3.0 provides users with the ability to control their information. However, maintaining the distributed Internet over a vast number of nodes presents significant challenges. In this context, 5G technologies play a pivotal role in deploying decentralization with low latency and unparalleled bandwidth perspectives. Specifically, the ultra-high speed connectivity from 5G promises a deployment of a decentralized Internet web, which operates through a network of devices, preventing data control by any single server~\cite{zeeve2023web3}. Furthermore, the specifications of 5G including URLLC, eMBB, and mMTC ensure continual peer-to-peer communications which is the most important in Web 3.0. Taking beamforming technology as an example, advanced signal processing methods are applied to minimize the interference between peer-to-peer nodes, therefore improving the performance of data transmission in a cluster of decentralized communication nodes~\cite{cheng2012beam}.  

\subsubsection{Diversity and efficient data transmission}
Data transmission plays a significant role in the Web 3.0 environment in which various technologies are utilized. As Web 3.0 requires the transmission of various data types across layers that are directly linked to different technologies, 5G can offer significant benefits in this field~\cite{ray2023web3}. By offering low latency and high bandwidth from advanced wireless technologies like massive MIMO, URLLC and beamforming, 5G enables Web 3.0 to operate more efficiently and ensures fluency in the data streams, thereby increasing the experience of users.

\subsubsection{Immersive and interactive web content}
Web content is another crucial aspect of Web 3.0 which enables augmented reality, virtual reality and 3D simulations to increase user experience. The challenge in deploying such rich and complex technologies in real time requires reliable data transfer, and 5G technologies can solve that problem. According to Insidetelecom~\cite{zeeve2023web3}, the necessity for rapid internet in developing virtual technologies in Web 3.0, including the Metaverse, makes 5G technologies essential. Thanks to the ultra-low latency and massive data throughput of 5G, users can access virtual spaces with high quality~\cite{wang2023low}. Besides, by processing data near the source, edge computing can improve Web 3.0 usage with the ability to make real-time interactions with no delay. Additionally, the network slicing method of 5G can also customize the network allocations~\cite{liu2023slicing4meta}. Therefore, the demanding web content becomes a priority with efficient resources.

\subsubsection{Massive connectivity}
5G technologies play a significant role in empowering the connectivity of Web 3.0. It enables Web 3.0 to create a more decentralized, intelligent and diverse environment with the advanced methods of 5G as the backbones. Leveraging such technologies as Massive MIMO, edge computing and network slicing allows the deployment of a massive number of devices from IoT sensors to autonomous vehicles~\cite{alex2019mimo}. These devices can operate with high-data processing and provide real-time feedback to users with extremely low latency. This opens many new use cases like smart cities, remote surgery, intelligent transportation, online learning and live streaming, which can make Web 3.0 become more immersive. 5G technologies not only improve IoT connectivity but also foster new possibilities in DApps, augmented reality and virtual reality. As a result, the integration of 5G connectivity and edge computing can enable real-time processing for a million devices and support data-intensive applications that leverage augmented and virtual reality~\cite{zeeve2023web3}. This leads to the enhancement of user experiences with swift connections and consistent machine-to-machine interaction.

\begin{figure*}[!]
    \centering
    \includegraphics[width=0.8\textwidth]{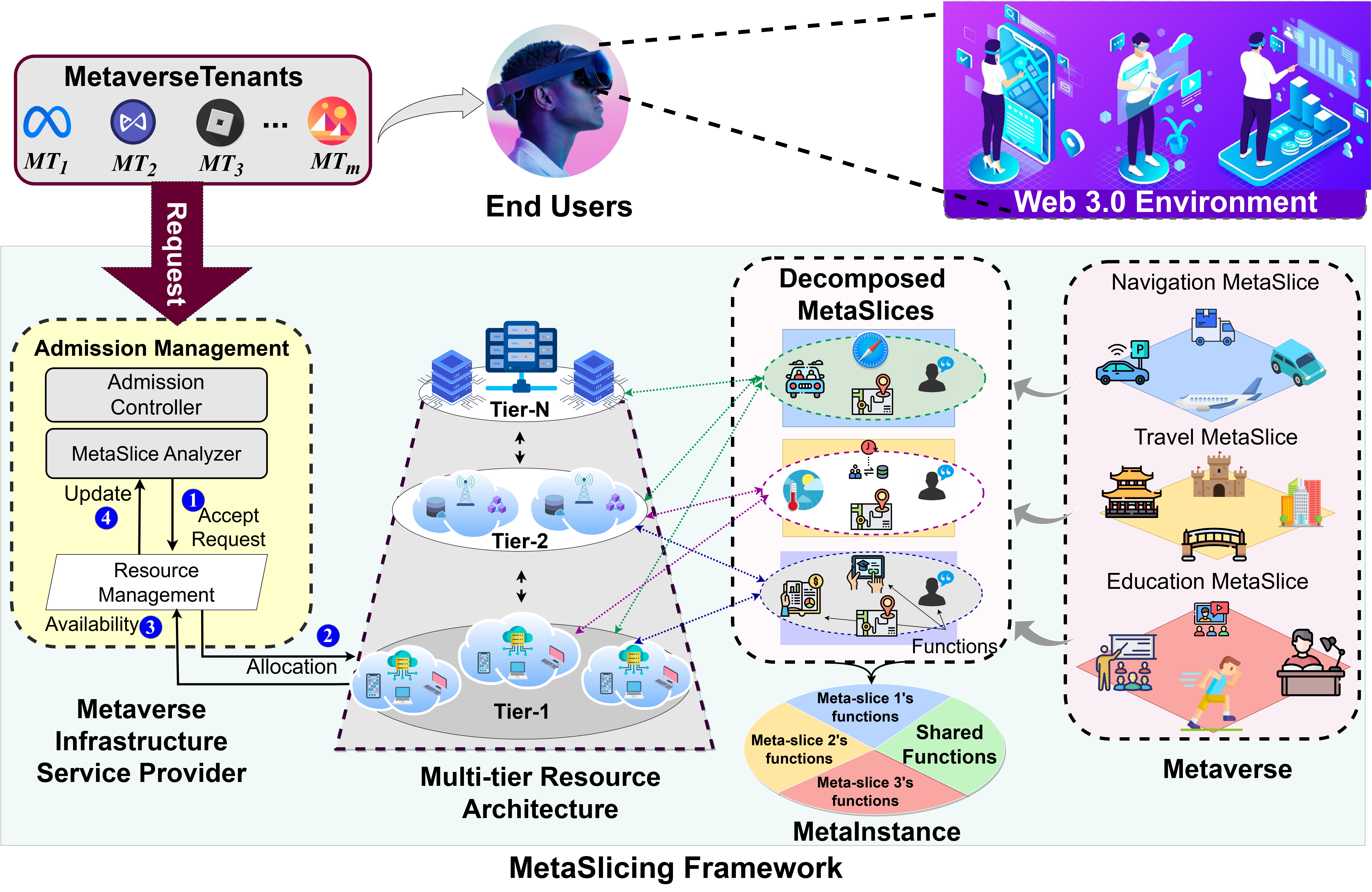}
    \caption{The proposed architecture in~\cite{nam2023slicing} aims to provide intelligence Metaverse application management based on the decomposition of each application and requests processing by deep reinforcement learning.}
    \label{fig:metaslicing}
\end{figure*}

\subsection{Recent Research on 5G Technologies for Web 3.0}

\subsubsection{Performance improvement for Web 3.0}

5G is a cutting-edge wireless technology that enables faster and more reliable communication for various applications. It has three critical use cases: URLLC, eMBB and mMTC~\cite{imt2015vis}. These use cases aim to improve the quality of experience (QoE) for users by reducing latency and increasing bandwidth. Therefore, 5G can enhance the performance of many innovative technologies, including Web 3.0. By integrating 5G and Web 3.0, the web can become more efficient and responsive, offering a superior user experience. 

As Web 3.0 offers user-centric, intelligent and autonomous service, it is expected to have communication and computation capabilities at the communication and network layers. This poses substantial challenges in achieving ubiquitous and low-latency communication given the heterogeneity and dynamics of current networks. To solve this problem, the authors in~\cite{wang2023low} illustrated an intelligent offloading communication. The deep reinforcement learning is adopted to explore the optimal data transmission between digital twins and base stations. The digital twins obtain the real-time statuses of the base stations to estimate the reliability and latency. The status of base stations that are satisfied for latency requirements is the input of a Deep Q-Network to explore and execute the offloading decision. The numerical results showed that the proposed architecture could provide successful offloading probability via different numbers of users, thereby improving the QoE of Web 3.0 with low latency communication.

Moreover, Metaverse is one of the exemplary applications of Web 3.0, which offers users interactive and immersive 3D experiences. However, it faces significant challenges in resource allocation, which requires enormous resources, including computing, storage and network resources, to support extended reality and ensure reliable and low-latency communication. Therefore, the authors in~\cite{nam2023slicing} proposed a novel resource allocation framework for Metaverse called MetaSlicing, which is illustrated in Fig.~\ref{fig:metaslicing}. In particular, the Metaverse applications are first decomposed into multiple functions, each of which can be independently initialized and positioned at a different computing tier to utilize all the network resources. The applications with common functions are then grouped into clusters called MetaInstances, allowing these functions to be shared across the layers. After that, a semi-Markov decision process integrated with a deep reinforcement learning technique is proposed in the service provider to find an optimal admission control policy for the Metaverse, thereby satisfying the high dynamic of resource requests and real-time response. The authors also highlighted that the proposed MetaSlicing can improve the performance of a Metaverse system, paving the way for dynamic resource allocation and interoperability between different MetaSlices in Metaverse, thereby boosting efficiency in Web 3.0.

In the Web 3.0 ecosystem, conventional end-to-end system is not effective due to the complexity of data processing within a decentralized network. This complexity comes from its reliance on information-centric networking and peer-to-peer data storage in which data is content-addressed. Therefore, the implementation of Named Data Network Functions (NDNF) is presented in~\cite{fang2023implementing}, enabling virtual network functions on data requests in NDN, which can be potentially applied to Web 3.0. In NDNF, network functions are encoded as part of the requested content name, and NDN mechanisms are applied to ensure network function execution in the node during data transmission. The authors showed a comprehensive simulation of NDNF with various network function configurations in a gateway router. The simulation results indicated the feasibility and practicality of NDNF not only in small-scale networks but also in complex ones with multiple request and network function services, thereby allowing the deployment of Web 3.0 with an information-centric and decentralized web environment.

\subsubsection{Semantic communications for Web 3.0 using 5G and beyond}

Semantic communication is a new paradigm of wireless communication that focuses on the transmission of the meaning (or the semantic information) of the source rather than the exact bits or symbols~\cite{luo2022sem}. It aims to overcome the limitations of the Shannon theory, which is based on the physical capacity of the channel and the accuracy of the received signal. In the context of Metaverse in Web 3.0, maintaining 3D data transmission between the physical environment and the digital world seamlessly throws a notable challenge~\cite{ning2023survey}. In this scenario, semantic communication becomes a solution by employing deep learning-based end-to-end communication and natural language processing techniques to enable the transmission of the semantic content of messages~\cite{xie2021sem}. This approach eliminates the need to transfer raw 3D data in Metaverse services, leading to efficient data transmission with resource allocation and low latency communication within Web 3.0, thereby improving the experience of users. Hence, applying 5G and advanced wireless technologies to enable semantic communication in Web 3.0 is essential, thereby bridging the gap between user interaction and digital feedback.

The authors in~\cite{hong2023semantic} proposed a semantic transmission framework to transfer information from the physical world to the Metaverse which can be illustrated in the Web 3.0 environment. The transmission cost and data storage can be improved by utilizing semantic space in both transmitter and receiver, owing to the migration in transmitted sensing information. In the considered framework, the transmitter focuses on semantic encoding and transferring information following the calculation of channel frequency response. Meanwhile, the receiver extracts semantic features from the incoming signal in terms of channel decoding. Additionally, the receiver can also restore the complete channel frequency power to facilitate other Metaverse services. Besides, a theory reward mechanism is provided to incentivize the transmitter to send data with high frequency, thereby increasing the user experience in the Metaverse. The numerical results showed that the proposed framework can reduce the amount of transmitted data and effectively boost the transferred data in high frequency.

Alternatively, the authors in~\cite{luong2023edge} deployed semantic communication in virtual service providers (VSPs) in the context of Metaverse to reduce the collected data and offloading cost from the physical devices. In the suggested system, semantic communication is applied in UAVs, which is deployed by VSPs to reduce data traffic and computing costs. By leveraging the convolutional neural network (CNN) in the RelTR technique, a CNN is combined with a feature encoder and entity decoder to generate the feature context and entity representation of the images captured by the UAV. It employs a Triplet Decoder that utilizes attention mechanisms to classify and relate subjects, objects, and their predictive, effectively capturing the relationships within the image. These relationships are structured into semantic triplets, including subject, predicate and object, via feedforward neural networks, translating raw data from the physical world into a semantic symbol, therefore mitigating data transmission size from UAV to digital twins. The authors demonstrated that applying semantic communication can reduce the offloading cost in the service providers, thereby enabling the deployment of immersive 3D content in a Web 3.0 environment.

\subsubsection{Security and privacy in Web 3.0}
The integration of 5G into Web 3.0, while offering advanced data transmission and connectivity, also introduces significant security and privacy challenges that are essential for guaranteeing uninterrupted and secure massive data transmission in the web environment. In the context of Web 3.0, it involves many technologies and intermediate systems, and thus it has many open interfaces which are vulnerable to diverse attacks. For example, IoT devices at the physical layer can be disrupted by common communication and network attacks such as DDoS, Sybil, Replay, etc, thereby decreasing the performance of Web 3.0~\cite{mar2019sec}. In addition, emerging applications like Metaverse also present significant security risks which can negatively impact Web 3.0. For instance, by virtualizing real-world environments instantaneously, users can face malicious actions including virtual stalking and spying, illegal tracking locations, digital asset hijacking, and deepfake events~\cite{wang2023meta}. Consequently, security considerations are crucial in the integration of 5G with Web 3.0.

Recently, authors in~\cite{liu2023secure} proposed a mechanism to ensure security and privacy in Web 3.0 in the context of IoV. In the considered system, an anomaly detection algorithm for edge server placement is built based on zero-trust security and a non-cooperative game model, thereby preventing cyber-attacks from outliers to cloud servers in the IoV-based cloud edge computing in Web 3.0. All tasks and users' requests are considered untrusted in the IoV environment and must be verified before accessing the system. As the anomalies are assumed to increase the workload and generate computational failure, the security mechanism is based on the data size of tasks. The tasks are decided to be insecure if their data size is less than the multiply of the coefficient and average of data sizes. The simulation results showed that the proposed algorithm can effectively ensure the security between the edge server and base stations, maintaining undisrupted and low latency communication, thereby improving the performance of Web 3.0 empowered by IoV.   

\textit{Summary}: 5G technologies play an essential role in the development of Web 3.0 by providing ultra-low latency and highly reliable internet connectivity. It facilitates the seamless communication of advanced Web 3.0 applications, e.g., AR/VR and IoT devices, which require massive data transmission and real-time processes. The high data transfer rates and high bandwidth utilization of 5G ensure that Web 3.0 with intensive data can run smoothly, fostering a more interconnected and responsive digital environment. With the enhanced network capabilities, 5G is a fundamental infrastructure supporting the decentralized, user-centric, and autonomous vision of Web 3.0, enabling more dynamic and immersive online experiences. A summary of the 5G technologies approaches for Web 3.0 is presented in Table~\ref{table_5G}.

\begin{table*}
	\caption{Summary of Integrating 5G Technologies in Web 3.0}
	\label{table_5G}
	\begin{centering}
		\begin{tabular}{|>{\raggedright\arraybackslash}m{1.0cm}|>{\raggedright\arraybackslash}m{1.7cm}|>{\raggedright\arraybackslash}m{3.0cm}|> {\raggedright\arraybackslash}m{4.0cm}|>{\raggedright\arraybackslash}m{6.2cm}|}
			\hline 
			\multicolumn{1}{|>{\centering\arraybackslash}m{1.0cm}|}{\multirow{1}{*}{\textbf{Ref.}}} & 
            \multicolumn{1}{>{\centering\arraybackslash}m{1.5cm}|}{\multirow{1}{*}{\textbf{Application}}} &
            \multicolumn{1}{>{\centering\arraybackslash}m{2.9cm}|}{\multirow{1}{*}{\textbf{Research problem}}} &
            \multicolumn{1}{>{\centering\arraybackslash}m{4.0cm}|}{\multirow{1}{*}{\textbf{Solution}}} &
            \multicolumn{1}{>{\centering\arraybackslash}m{6.2cm}|}{\multirow{1}{*}{\textbf{Result}}}\\
			\hline 
			\hline

          \cite{fang2023implementing} & General Web 3.0 application & Addressing the information-centric and content-addressed data on Web 3.0 & Network functions for content of requested data and named data network for data transmission & The proposed method can successfully process multiple request and services with data /name included in transmitted data. 	\\	
          \cline{1-5} 

     \cite{wang2023low} & General Web 3.0 application & Dynamic perception and intelligent decision-making & Digital twins for estimating latency and deep reinforcement learning for tasks offloading decision in Web 3.0  & Enhance low latency communication with low delay (less than 0.2 ms) and efficient offloading decisions (up to 40\%).  \\ 	
      \cline{1-5} 

        \cite{nam2023slicing} & Metaverse & Resource allocation  & Multi-tier computing and network function for network resource and deep reinforcement learning for admission control & Enhance resource management with high reward compared with existing works (up to 80\% greater). 	\\	
           \cline{1-5}

       \cite{hong2023semantic} & Metaverse & Efficiency sensing information transmission from IoT devices & Semantic transmission for both transmitter and receiver &  Enhance data transmission after semantic encoding with the reduction in data amount up to 27.87\%.   \\ 	
        \cline{1-5}

        \cite{luong2023edge} & Metaverse & Distribute and trading computing resource & DRL technique is used to maximize the revenue for the service provider and semantic communication is utilized to reduce the collected data and offloading cost  & Efficiently enhance offloading tasks by translating raw images with 3.59 Mbytes to semantic symbols with up to 0.65 Mbytes. 	\\	
        \cline{1-5} 
   	
     \cite{liu2023secure} & Internet of Vehicles & Detect and prevent malicious computing offloading tasks &  Zero-trust security for anomaly detection and a game model is used to enhance the security and optimize system performance  & The proposed method can ensure security while maintaining low latency in IoV communication.   	\\	
    \cline{1-5}  
        
			\hline                                                       
	  	\end{tabular}
		\par\end{centering}
\end{table*}

\section{Blockchain Technology for Web 3.0}
\label{sec:blockchain}

This section delves into the fundamental concepts of blockchain technology, providing its essential components and mechanisms. Following that, we investigate the important roles of blockchain technology in Web 3.0. Furthermore, we examine the most recent developments and technological advances in blockchain technology, highlighting significant studies and findings that are contributing to the development of blockchain into the foundation of Web 3.0.

\subsection{Blockchain Fundamental}

Blockchain technology represents a fundamental aspect of Web 3.0 that has gained significant attention as a groundbreaking concept. In the conventional database systems that centralize data storage management, making them susceptible to attacks and data breaches~\cite{korkmaz2022alder}, blockchain functions as a decentralized database spread across multiple decentralized nodes~\cite{sunny2022systematic}. The fundamental idea behind blockchain is to establish a transparent, secure, and unalterable system for storing and verifying transactions, making it highly challenging for any single entity to modify or tamper with the data~\cite{al2019blockchain} \cite{fan2020performance}. A blockchain system encompasses three key attributes:

 \begin{itemize}
 
 \item \textit{Decentralization}: Decentralization plays a key role in blockchains, as it distributes information across a network of nodes, ensuring that no single entity has full control~\cite{ray2023web3}. This enhances security and resilience, while also eliminating any single points of failure or vulnerability~\cite{casino2019systematic}. By leveraging blockchain technology, we can build decentralized communication networks that are resistant to censorship and surveillance \cite{ray2023web3}. This empowers users with increased privacy and control over their data \cite{yang2019integrated}. 
 
  \item \textit{Transparency}: Transparency is another essential element of blockchain that shapes a decentralized and trustworthy environment where users can participate with confidence, knowing that their interactions are secure, auditable, and accountable. The blockchain ledger is accessible to the public, allowing anyone to assess and verify network transactions without a third party~\cite{zhou2021building}. Users independently check the authenticity and integrity of transactions, ensuring that data is correct and tamper-proof~\cite{khalid2023comprehensive}.

 \item \textit{Immutability}: Immutability is another important aspect of blockchain, that ensures that the data stored on the blockchain cannot be changed or erased by anyone~\cite{zhou2021building}. Blockchain creates digital signatures that link each block. This guarantees that any modifications to a transaction must be granted by a majority of nodes, making the system very resistant to fraud and manipulation \cite{monrat2019survey}. As a result, immutability can ensure that the blockchain is secure, transparent, and trustworthy, as no one can manipulate or falsify the data~\cite{gai2019permissioned}. Furthermore, immutability also improves efficiency and reduces the costs of various processes that involve data recording and sharing, such as contracts, transactions, and audits by using hashing, digital signatures, and consensus mechanisms~\cite{monrat2019survey}.

\end{itemize}
 
Consensus mechanisms play a crucial role in attaining agreement among participants in a decentralized network regarding the validity and order of transactions. They hold paramount importance in upholding the integrity and security of the blockchain. Consensus mechanisms typically resolve mathematical problems or designate trusted nodes to verify transactions~\cite{guan2018privacy}. Once consensus is achieved, the transactions are added to the blockchain, ensuring that all participants possess a consistent and mutually agreed-upon update of the ledger. Consensus mechanism prevents fraud and double-spending attacks and ensures a shared understanding of the blockchain state~\cite{korkmaz2022alder}, \cite{tanwar2022blockchain}, \cite{nguyen2023fedchain}.

Proof of Work (PoW) and Proof of Stake (PoS) are the most popular consensus mechanisms in blockchains. PoW requires miners to solve complex puzzles, ensuring the integrity and security of the blockchain through computational power~\cite{miller2015nonoutsourceable}. However, it is energy-intensive. PoS, on the other hand, selects validators based on their stakes (i.e., tokens) in the network, making it more energy-efficient~\cite{yan2022analysis}. Both have strengths and weaknesses, and ongoing research explores hybrid approaches to improve scalability and sustainability. The decision between PoW and PoS is determined by each blockchain network's unique demands and goals.

In addition to these mechanisms, there are also some other mechanisms, such as Proof-of-Elapsed Time (PoET), Proof-of-Authority (PoA), Proof-of-Authentication (PoAh), Proof-of-Property (PoP), Proof-of-Capacity (PoC), Proof-of-Burn (PoB) and Proof-of-Weight (PoWeight) \cite{zarrin2021blockchain}. PoET is a consensus mechanism that utilizes a randomized waiting period for validators in a blockchain network to compete for the creation of the next block \cite{chen2017security}. PoAh eliminates the reverse hashing function in favor of a lightweight, resources block validation process~\cite{puthal2020poah}. PoA is a more effective consensus technique that solves the concerns of PoW's high latency, low transaction rate, and power consumption~\cite{liu2019mdp}. Po-Activity combines PoW and PoS to support authentic transactions and consensus among miners~\cite{liu2019fork}. PoP offers ``proof'' for blockchain information structure features, which uses less energy, and permits nodes to reduce the quantity of information required for each transaction~\cite{zarrin2021blockchain}. PoB, PoC, PoWeight, and PoL, each introduce a new approach to achieve consensus. Each of these mechanisms has its own strengths and weaknesses, and the choice of the consensus mechanism depends on the specific requirements and goals of the blockchain network.

\subsection{Roles of Blockchain in Web 3.0}

The emergence of Web 3.0, with its notable focus on decentralization, is a key driver in shaping an open and digital landscape. It gives users more power to control their data~\cite{zarrin2021blockchain}, \cite{ray2023web3}. Unlike conventional centralized applications where user data is controlled by a centralized datastore, blockchain technology enables secure, trustless DApps that are less susceptible to vulnerabilities and can execute tasks automatically using smart contracts~\cite{hewa2021survey}. Embracing blockchain in Web 3.0 yields several advantages, including heightened security, scalability, transparency, and efficiency \cite{Liu2022MakeWC}. By incorporating blockchain technology, the risk of information leakage and threats can be reduced, while transaction costs are minimized, and transaction speeds are increased~\cite{sunny2022systematic}. The roles of blockchain technology in Web 3.0 are presented in the following subsections.

\subsubsection{Decentralization of data and services}
Decentralization has been transforming how online systems work by spreading out data and applications across a network. The primary goal of Web 3.0 is to build an intelligent network system that will not be fully relied on by a single authority. Blockchain technology plays a significant role in achieving decentralization thanks to its unique consensus mechanism and distributed ledger system~\cite{zarrin2021blockchain}. In a blockchain network, numerous nodes collaborate in validating and verifying transactions. A consensus protocol ensures that everyone agrees on the current state of the ledger~\cite{wang2019survey}. This consensus is reached without the need of a central authority because each node autonomously maintains a copy of the entire blockchain. As a result, there is no single point of control or failure, and data is securely distributed across the network~\cite{singh2020sidechain}. This decentralized approach eliminates the risk of a single entity manipulating the data, thereby promoting a more equitable and resilient system in Web 3.0.

\subsubsection{Security and privacy}
Security and privacy are another crucial part of the sustainable development of Web 3.0.  The structure and features of blockchain technology bring a significant boost to security and privacy within Web 3.0. It provides users more power to manage their own personal sensitive data, digital assets, and identities.  Data is securely kept in an immutable ledger on a blockchain network, guaranteeing that no one can access or alter it without authorization~\cite{tanwar2022blockchain}. Each transaction is encrypted and connected to the one previous to it, producing an unbroken data chain~\cite{chen2019toward}. This design makes exceedingly difficult for malicious actors to alter the information subsequently and minimizes the potential for malicious attacks such as personal information disclosure and transaction records.

\subsubsection{Tokenization}
Tokenization is another significant concept in the Web 3.0 ecosystem, which uses blockchain technology to transform asset representation and exchanges. It is the technique of encoding physical or digital assets into unique tokens throughout a blockchain network~\cite{zhou2019dlattice}. These tokens are used to represent ownership, values, and other advantages and features~\cite{lin2023unified}. It also supports transparency and efficient asset transfer by removing the need to use middlemen and related expenses. Moreover, it is the key element of the Web 3.0 ecosystem and has the potential to transform ownership, value, and trade to the digital world.

\subsubsection{Digital Identity}
Digital identity plays a vital role in ensuring secure and trusted interactions in Web 3.0. Online identity management faces serious challenges such as security and privacy. Blockchain technology, however, could be utilized to address these problems by offering a decentralized identifying system~\cite{wang2023linking}. It enables users to keep their identities confidential and decentralized.  Blockchain technology provides a cryptographic key that allows web users to access and authorize themselves securely. By utilizing blockchain technology, it can be possible to establish the foundation for a new era of digital identity management for Web 3.0. In such a way, the web users will have complete control over their online interactions as well as their digital identities~\cite{furfaro2019infrastructure}.

\subsubsection{Transparency and immutability}
Transparency and immutability are among the most important foundations of Web 3.0. Blockchain technology plays a pivotal role in upholding these principles. Transparency ensures openness and accessibility of data by providing a shared record of all transactions and actions, allowing users to independently verify data without relying on intermediaries~\cite{zarrin2021blockchain}. This ensures the integrity and provenance of unique digital assets or data secured. On the other hand, immutability guarantees that once the data is recorded in the blockchain, it cannot be modified or changed. Blockchain ensures this by using decentralized consensus methods and cryptographic hashing, making it nearly impossible to manipulate data~\cite{tanwar2022blockchain}. Together, transparency and immutability bring a secure and trustworthy environment digital landscape in Web 3.0.

\subsection{Recent Research on the Development of Blockchain Technology for Web 3.0}
 \subsubsection{Smart contracts for transactions management in Web 3.0}
Smart contracts are considered as a vital component of the Web 3.0 ecosystem which eliminates traditional contracts. It is a self-executing contract encoded in computer code that eliminates the need for middlemen~\cite{ren2023building}. When the predefined circumstances are satisfied, these contracts automatically execute conditions between parties and are stored on a blockchain network~\cite{zou2019smart}. Smart contract eliminates middlemen, reducing expenses, and ensuring secure transactions on the blockchain network~\cite{wu2022review}. It provides control over users' sensitive data and privacy in Web 3.0 by offering a secure and transparent way of storing and executing agreements between devices, and ensuring that information is only distributed to legitimate parties~\cite{ren2023building}. This keeps users' data private and secure, reducing the risk of illegal access or manipulation.

The authors in \cite{salim2021blockchain} proposed an effective privacy framework that enables IoT devices (i.e., participants) to share data with a Social media 3.0 (SMW3) service provider while maintaining their privacy. In the considered system, the task requirements from the SMW3 service providers are sent to the blockchain system via smart contracts.  These smart contracts ensure the secure and automated execution of the task requirements. The requirements are then broadcast to the participants (i.e., IoT devices). If an IoT device has relevant data to a request, it accepts the smart contract and initiates the training process. Note that the participants only need to share their trained models (i.e., knowledge learned from their data via the training process) instead of sharing their raw data with the service providers. The authors showed that the proposed framework can protect the privacy of participants with various criteria for utility and privacy. Alternatively, the proposed framework can facilitate secured and automatic transactions among, participants, SMW3 service providers, and blockchain networks, facilitated by the implementation of smart contracts.

\subsubsection{Decentralized data storage management in Web 3.0}

Decentralized storage is an essential component of Web 3.0 which offers a novel alternative to conventional centralized data storage systems. In centralized storage, data stored on servers, which are controlled by a single entity, sometimes encounter storage and privacy problems~\cite{jia2023cross}. Thankfully, decentralized storage can address this problem by offering secure data storage workflows that dramatically speed up the management of data~\cite{zarrin2021blockchain}. By using cutting-edge blockchain technology, decentralized storage creates distributed networks of nodes that offer several advantages such as redundancy, enhanced security, and resistance to censorship~\cite{roberttechtarget}. This distributed decentralized nature makes it challenging for attackers to steal data, ensuring a more robust and tamper-resistant system~\cite{casino2019immutability}.  

\begin{figure*}[!]
    \centering
    \includegraphics[width=0.8\textwidth]{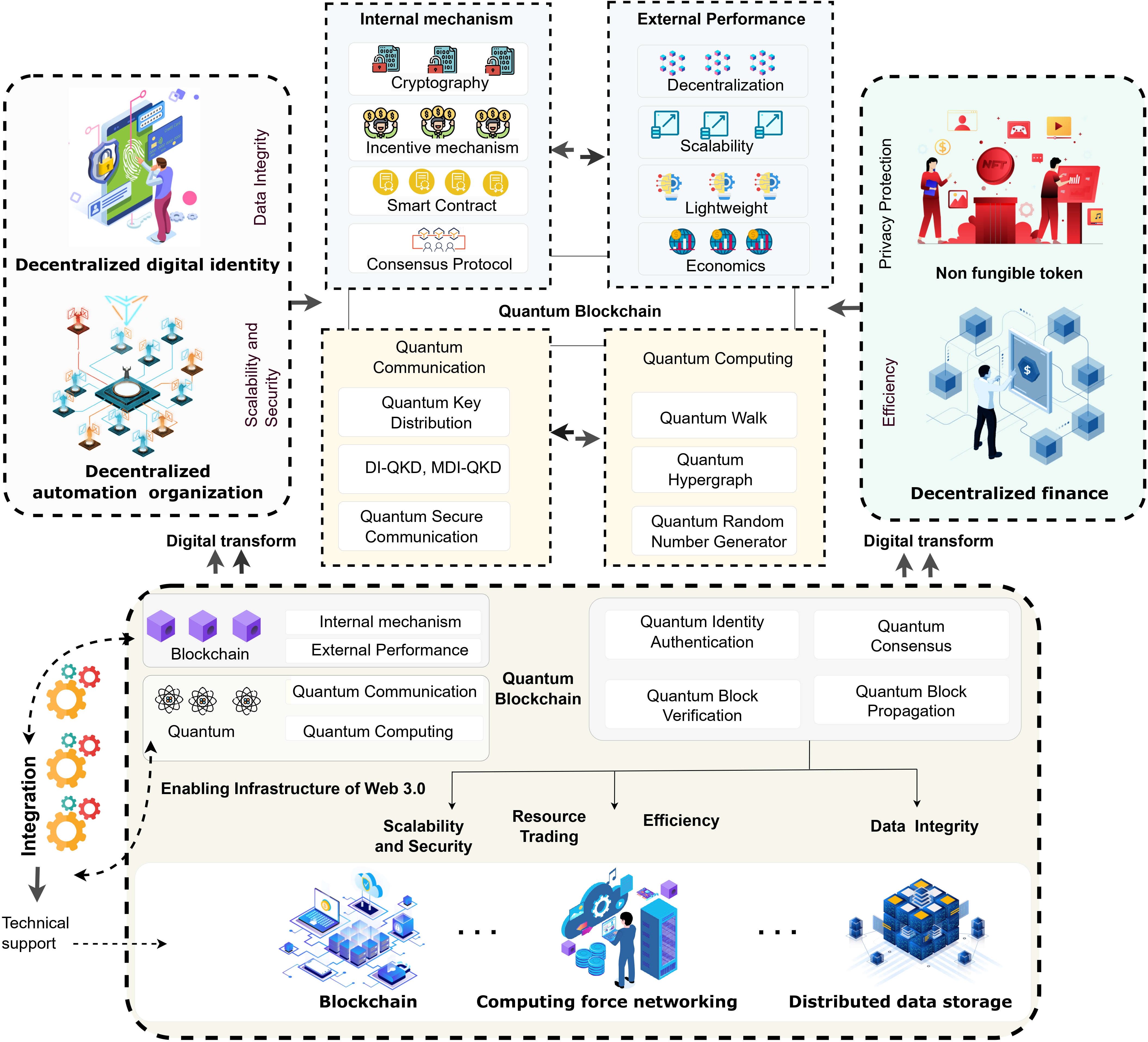}
    \caption{The proposed quantum blockchain-driven Web 3.0 framework adopted from~\cite{xu2023quantum}. The framework includes core infrastructure, quantum cryptography protocols, and services based on quantum blockchain technology in order to support a decentralized digital society in the quantum age.}
    \label{fig:blockchainweb3}
\end{figure*}

\begin{table*}
	\caption{Summary of blockchain-based approaches for Web 3.0}
	\label{table_blockchainapproach}
	\begin{centering}
		\begin{tabular}{|>{\raggedright\arraybackslash}m{0.8cm}|>{\raggedright\arraybackslash}m{1.7cm}|>{\raggedright\arraybackslash}m{4.0cm}|> {\raggedright\arraybackslash}m{4.0cm}|>{\raggedright\arraybackslash}m{5.0cm}|}
			\hline 
		\multicolumn{1}{|>{\centering\arraybackslash}m{0.8cm}|}{\multirow{1}{*}{\textbf{Ref.}}} & 
            \multicolumn{1}{>{\centering\arraybackslash}m{1.5cm}|}{\multirow{1}{*}{\textbf{Application}}} &
            \multicolumn{1}{>{\centering\arraybackslash}m{3.9cm}|}{\multirow{1}{*}{\textbf{Research problem}}} &
            \multicolumn{1}{>{\centering\arraybackslash}m{3.9cm}|}{\multirow{1}{*}{\textbf{Solution}}} &
            \multicolumn{1}{>{\centering\arraybackslash}m{4.2cm}|}{\multirow{1}{*}{\textbf{Result}}}\\
			\hline 
			\hline

          \cite{salim2021blockchain} & Social media & Data privacy concerns on social media while sharing data & Smart contract to ensure data privacy-while data sharing & Enhance data privacy, integrity, and efficiency compared to existing work by 97.55\%  	\\	
          \cline{1-5} 

     \cite{drakatos2021triastore} & Smart city &  Efficient data sharing for training AI using IoT data in Web 3.0 & Store locally trained models as transactions. Participants receive currencies as rewards for their performance. A peacemaker entity ensures protocol compliance. & Enhance data quality, performance, and data storage security resulting in high accuracy (up to 95\%) and minimal learning loss (up to 10\%).  \\ 	
      \cline{1-5} 

        \cite{zhang2023ai} & Metaverse & Ensure privacy, data ownership, and decentralization for a storage system for Web 3.0 & Store data in blocks to ensure privacy, protect data ownership, and enable
        transactions without third parties. Smart contracts to secure data transactions and verification & Improve data privacy, data ownership, and Metaverse performance compared to the existing work for $360^o$ video streaming  	\\	
           \cline{1-5}

       \cite{wang2023linking} & User identity & Addressing user identity issues on Web 3.0 & Zero-knowledge proof for users' identity verification &  Reduce the amount of time required to verify users' identities effectively (up to 126 ms).   \\ 	
        \cline{1-5}
                  
        \cite{lin2023unified} & General applications & Addressing semantic verification in various Web 3.0 scenarios & Use edge servers to encrypt and verify data and smart contracts to record and access data  & Enhance semantic communication/verification, data processing, and efficiency compared to existing works\\	
        \cline{1-5}

       \cite{xu2023quantum} & NFT and payment security & Ensures security and privacy of users’ identity in quantum-based blockchain services &  Quantum signatures to verify user identities, quantum NFT tokens for assets ownership & Enhance the security and privacy of web 3.0 user identities\\	
       \cline{1-5} 
   	
     \cite{lin2023blockchain} & Finance & Addressing the gaps in blockchain and semantic communication efficiency and performance on Web 3.0. & Zero-knowledge proof for semantic verification and Stackelberg game for semantic trading  & Enhance data efficiency and performance in Web 3.0  	\\	
    \cline{1-5} 

       \cite{doe2023promoting} & Metaverse & Addressing Web 3.0 data processing costs and associated challenges & Contract-theoretic incentive mechanism to reduce service cost  & Enhance network's utility and reduce cost utility overall 54.52\% and 62.5\%, respectively		\\	
        \cline{1-5}
			\hline                                                       
	  	\end{tabular}
		\par\end{centering}
\end{table*}

In~\cite{drakatos2021triastore} the authors proposed blockchain-based decentralized storage in smart cities for Web 3.0. The proposed framework simplifies machine learning by using storage blocks to store the shared database and nodes to handle querying and updating the blockchain ledger. In particular, the users stream and share their locally trained models to decentralized nodes, which are then stored as transactions in the distributed ledger. The previous round's node winner creates blocks with transactions using a dual consensus protocol. Users participating in the blockchain procedure receive training currencies as rewards based on their performance. A peacemaker entity ensures protocol compliance and prevents fraudulent behavior. It claims coins as a reward for ensuring protocol correctness, and all these steps are recorded in decentralized storage.

A similar decentralized data storage for semantic verification/communication was proposed in \cite{lin2023unified}. In the considered system, users collect data using wearable devices from Web 3.0 applications. This data is sent to the system for encryption. Edge servers then encrypt the data based on request. The server searches for local or shared knowledge to identify elements involving storage, verification, and communication operations. The exchanged data is then actively recorded on the blockchain through smart contracts after being retained off-chain by edge servers. Physical channels are used to transmit the data and facilitate interaction between data producers and consumers to avoid overloading data improving interaction efficiency in Web 3.0. The authors showed that the proposed framework significantly enhanced semantic communication, data processing, and efficiency compared to the existing work.

Moreover, decentralized storage can be used for Web 3.0 Metaverse applications that provide reliable, secure, storage solutions for data transmitted within virtual worlds and immersive environments \cite{liu2023slicing4meta}. For example, the authors in \cite{zhang2023ai} proposed a decentralized storage system for Web 3.0 Metaverse applications. In the considered framework, physical world data, such as videos, are gathered through sensors and uploaded to the virtual space to construct digital avatars and environments. The virtual world interacts with user data from the physical world and provides feedback. The data is collected from the system and stored on decentralized storage to ensure privacy, protect data ownership, and enable transactions without third parties. Smart contracts are used to secure data transactions and verification within the blockchain network. The authors showed that the suggested framework boosts the productivity of Metaverse video transmission and effectively ensures secure data storage.

\subsubsection{Tokenization and digital asset management in Web 3.0}
 Tokenization and digital assets open up new opportunities for developing DApps and new business models in Web 3.0 by reflecting real-world assets (i.e., land, avatars, and virtual goods). It is the process of transforming real-world assets or rights into digital tokens on a blockchain network~\cite{zhou2019dlattice}. Users can easily purchase, sell, and trade these assets as tokens on a blockchain by eliminating the need for middlemen. It also verifies the ownership records, and transaction histories are securely preserved and validated. Traditional information systems commonly utilize centralized data and require intermediaries to manage transactions. This can lead to high transaction costs, lack of transparency, and limited access to rewards opportunities~\cite{cao2022decentralized}. However, tokenization can address these issues by providing a standardized way of representing information and data organization in the virtual world and an incentive for user engagement~\cite{wang2023meta}.
 
 Recently, a token-based semantic exchange framework for Web 3.0 was presented in~\cite{lin2023blockchain}. The authors used NFTs, zero-knowledge-proof, and semantic extraction to enhance the effectiveness of Web 3.0's interactions. In the considered system, smart contracts utilize tokens to play a vital role in facilitating data exchange. Users effectively upload data to off-chain storage using the interplanetary file system (IPFS) to access metadata. Smart contracts and blockchain technology are utilized to semantically store data, generate NFT tokens, and distribute them to the network. This enables the exchange of off-chain semantic data and the profit potential. Subsequently, a Stackelberg game is employed to optimize semantic pricing strategies. The proposed solution offers significant improvements to trust, transparency, and efficiency in Web 3.0. The framework is examined through an urban planning case study and proven to be efficient and effective.

A similar token-based incentive mechanism for the Metaverse Web 3.0 application was proposed in~\cite{doe2023promoting}. However, in this work, the authors focused more on the incentives entities driving the development of the blockchain network. In particular, the blockchain network evaluates users' ability to generate incentive resources and provides contract opportunities to the users. Once the users accept the offer, the smart contract analyzes user data, categorizes incentives, and distributes them to the users based on their performance. Furthermore, a reward classification system guides the selection and terms of trade for the offered rewards, establishing a connection between optimizing funding incentives and user reward determination in the blockchain network. The authors showed that the proposed framework can enhance security, and productivity and reduce costs in the interaction between users and the blockchain network in Web 3.0.

 \subsubsection{Decentralized identity management in Web 3.0}

Decentralized identification is another critical component of Web 3.0, which uses blockchain technology to provide a secure decentralized user-controlled identity ecosystem. In the conventional identification system, maintaining and protecting users' digital identities online is a critical issue leading to serious concerns about privacy and data breaches~\cite{jia2023cross}. However, these challenges can be mitigated using decentralized identity to ensure data privacy and security. The users can manage their individual digital identities using decentralized identities without eliminating third-party service providers~\cite{ray2023web3}. It enables the users to create and maintain their digital identities, as well as control their data and privacy. These identities are protected using secure keys and can be potentially used to identify users across platforms and services with accessible identity verification and authentication~\cite{avellaneda2019decentralized}.

  In~\cite{wang2023linking} the authors proposed a decentralized identity framework for Web 3.0. The proposed framework establishes a one-to-one link between decentralized identity accounts (i.e., souls) and users, effectively representing human interactions and ensuring user privacy and anonymity. In particular, the considered framework verifies users using facial recognition and then authenticates users with zero-knowledge proof. The authors utilized a consensus mechanism to encode facial recognition results and employ linkable ring signatures for user-account mappings within decentralized identity. The system is completely decentralized, relies on smart contracts for verification, and is tested on a practical blockchain network. The result illustrated that the proposed framework, incorporating decentralized identity, offers a privacy-preserving solution enabling users to maintain accountability for their accounts. A similar decentralized identity concept is proposed in~\cite{xu2023quantum}, which is illustrated in Fig.~\ref{fig:blockchainweb3}. The proposed framework comprises infrastructure, cryptography protocols, NFTs tokens, and quantum blockchain services. In particular, in the considered system, data is stored in distributed nodes, while cryptography ensures the security and privacy of users' identities. Users establish ownership assets through digital signing and NFT tokens. Quantum signatures verify user identities, enabling participation in information sharing and earning rewards using quantum blockchain tokens. The authors demonstrated that the framework can effectively deliver both anonymity and sustainability for the future decentralized society.

 \textit{Summary}: Blockchain technology will play a key role in the future development of Web 3.0 due to its outstanding features which are especially appropriate for Web 3.0. However, blockchain also faces several challenges when deploying in Web 3.0 platforms. As existing blockchain networks fail to effectively manage high transaction volumes, scalability is still an important challenge~\cite{zarrin2021blockchain} \cite{bhushan2021unification}. For wider adoption, interoperability between various blockchains and conventional systems is needed. To create a reliable and resilient blockchain ecosystem in the future, it is also necessary to resolve governing inconsistency and privacy issues~\cite{wang2022exploring}. Overcoming these challenges will be crucial to unlocking the full potential of blockchain in Web 3.0. A summary of the blockchain-based approaches for Web 3.0 applications is presented in Table~\ref{table_blockchainapproach}.

\section{The Semantic Web Technologies}
\label{sec:SWT}

The Semantic web technologies are an indispensable part for Web 3.0 development. Over twenty years ago, Sir Tim Berners-Lee presented the semantic web paradigm as a successor vision to Web 2.0~\cite{berners2001semantic}. This may lead to confusion between the semantic web and Web 3.0. They are sometimes used interchangeably; however, the semantic web is only a critical technology used in Web 3.0 (in addition to other enabling technologies like blockchain, AI and IoT). Instead of explicitly presenting information in specific textual and graphical forms, as in Web 1.0 and 2.0, the semantic web technologies aim to provide meaning/facts of information regardless of how it is presented by symbols. For example, the word ``jaguar'' can be represented by different symbols in different languages, as shown in Fig.~\ref{fig.SWW_meaning}. Moreover, the word can refer to various objects, such as animals, cars, and versions of the Macintosh operating system. This ability to express and understand multiple meanings of words allows the web to do more meaningful work. To that end, the semantic web technologies aim to provide intelligent data that machines can easily and automatically search, retrieve, and process ~\cite{bratt2007semantic}. In other words, the goal of the semantic web technologies is to make web data machine-readable and define how knowledge is represented so that humans and machines can work in cooperation.
	\begin{figure}[t]
    	\centering
    	\includegraphics[width=0.98\linewidth]{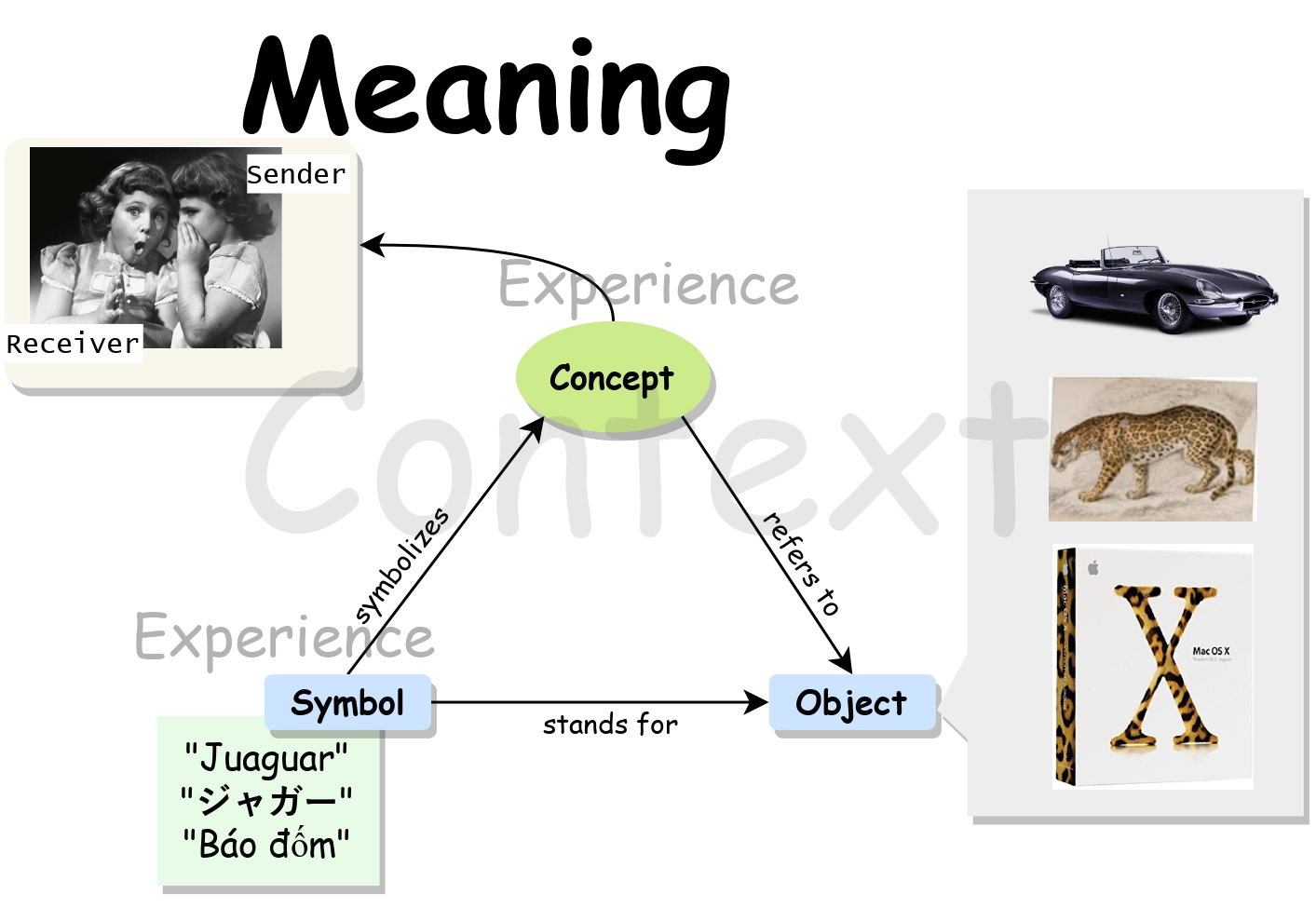}
    	\caption{The meaning of meaning: connection between language and thought~\cite{hitzler2009foundations}.}
    	\label{fig.SWW_meaning}
	\end{figure}
 
In practice, enabling automatic machine processing can be accomplished in two ways~\cite{bratt2007semantic}. 
	The first approach aims to build a smarter machine by teaching the machine to infer the meaning of data, possibly accomplished by AI-based methods, such as natural language and image recognition.
	The second approach, based on the semantic web technologies, focuses on building smarter data by expressing data and its meaning in a standard format that machines can automatically ``read''.
	In addition, the semantic web technologies also enable decentralizing data across the network. 
	Thus, the semantic webs are expected to transform the web from the ``web of documents'' e.g., Web 1.0 and Web 2.0 to the ``web of data''~\cite {w3cSemantic}.
	This section first presents fundamental technologies supporting the semantic web, called Semantic Web Technologies (SWT). 
	Then, the state-of-the-art in SWT is surveyed.
	Finally, the current challenges of SWT are discussed.

\subsection{Semantic Web Technology Stack}
Recall that semantic web technologies aim to provide machine-readable data so that machines can easily find and process data. In practice, this is a very challenging task. 
	For instance, the information \textit{``Beau Doe has a homepage http://www.example.com/beau''} can be easily understood by a human, but it is not an easy job for a machine.	
	To do so, semantic web technologies are developed to describe information in a structured and decentralized way. In general, the knowledge representation in the semantic web can be constructed from low to high levels~\cite{hitzler2009foundations}.  
	The first level defines an object, e.g., web resource using characters and syntax formats, e.g., Extensible Markup Language (XML). 
	This level corresponds to the two bottom layers of the Semantic Web stack, as shown in Fig.~\ref{fig.SWT_stack}. Then, the knowledge of the object (i.e., description) is given in the second level, corresponding to the third layer in Fig.~\ref{fig.SWT_stack}. 
	Finally, the third level, i.e., entire worlds, links knowledge among objects using various technologies from different types, such as Taxonomies, e.g., Resource Description Framework Schema, Ontologies, e.g., Web Ontology Language, and Rules, e.g., Rule Interchange Format. 
	Generally, the semantic web is developed based on this principle. 

The W3C, which is in charge of semantic web standardization, proposes the semantic web stack that groups the SWT into different layers, as shown in Fig.~\ref{fig.SWT_stack}. 
    In this stack, a layer uses services from its lower layer and provides services to its upper layer. 
    As such, each layer is built based on its lower layer, and it tends to be more complex than its lower layers.
		
\subsubsection{Web platform and syntax layers}
These first two layers build the foundations for semantic web technologies. 
    The web platform (the lowest layer) defines the web resources uniquely based on the Uniform Resource Identifier (URI)  and a character set, e.g., Unicode. 
    Then, the syntax layer defines the syntax/format of the data according to the Extensible Markup Language (XML).
		On the other hand, XML gives syntaxes for its upper layers, e.g., the Data Interchange, to describe information.		
		Note that the bottom two layers' technologies, e.g., URI, Unicode, and XML are well-defined in the current web and, without change, are the foundation for the semantic web.
		Thus, this stack shows that the semantic web aims to extend (not replace) the current web version.  
	\begin{figure}[t]
		\centering
		\includegraphics[width=0.99\linewidth]{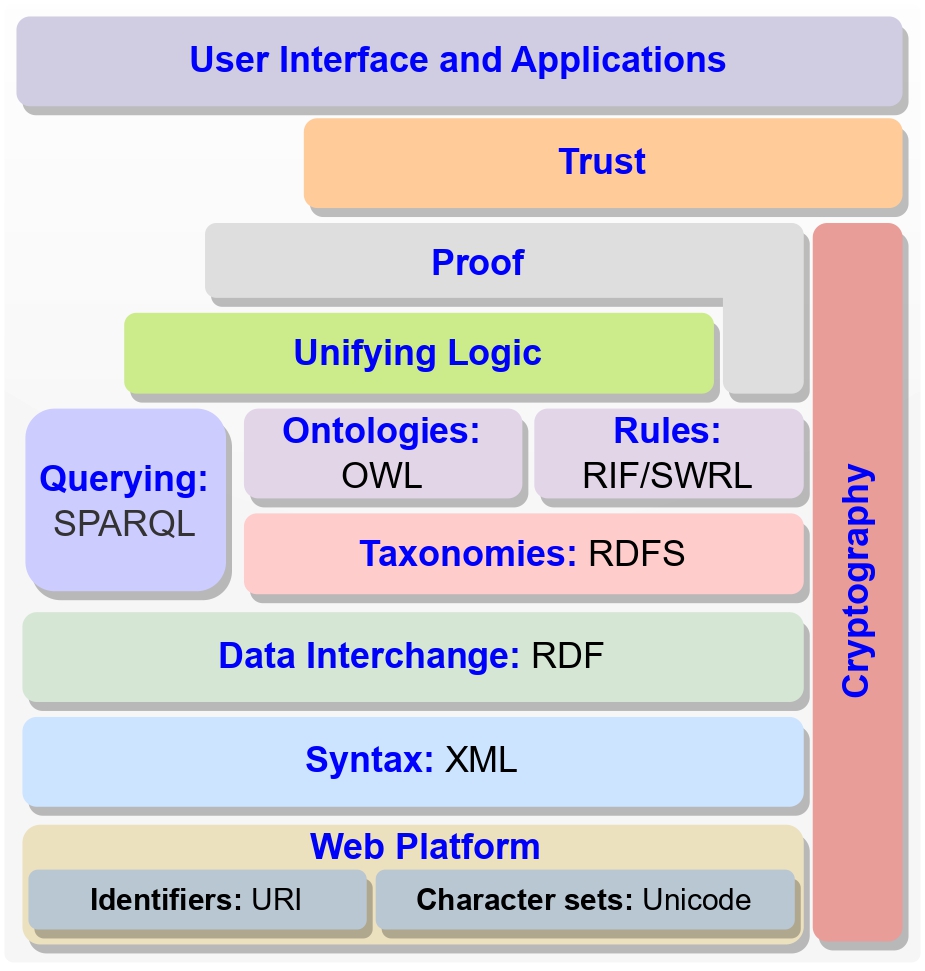}
		\caption{The Semantic Web stack~\cite{bratt2007semantic}.}
		\label{fig.SWT_stack}
	\end{figure}	

\subsubsection{Data interchange layer} 
Although XML is well-known for providing web document format, machines cannot understand it.
    As such, the Resource Description Framework (RDF) is proposed as a core data representation format for the semantic web~\cite{bratt2007semantic}. 		
    The RDF leverages URI to identify the resources (e.g., objects and subjects on the web) and describe them in terms of their predicates (i.e., properties and values).
    Thus, the RDF can be considered as a simple language to describe things, e.g., expressing relationships among web resources.	
                        
    In particular, RDF represents information by a triple, i.e., \textit{subject-predicate-object}, that forms a graph of data~\cite{w3cRDF}, as shown in Fig.~\ref{fig.rdf_triple_example}~(a). For example, the RDF triple of a statement \textit{``Beau Doe has a homepage www.example.com/beau''} is presented in Fig.~\ref{fig.rdf_triple_example}~(b). Here, all elements of this triple are resources with URI. Specifically, the first resource with URI \textit{www.example.org/beau/contact.rdf\#beaudoe} is an \textit{subject} that identifies Beau Doe.
    The second resource, i.e., ``has a homepage'', with URI \textit{www.xmlns.com/foaf/0.1/homepage} is the predicate pre-described in the XML format at the URI.
    The third resource is Beau Doe's homepage with URI \textit{www.example.org/beau}.     
    Note that in an RDF triple, all elements (i.e., subject, predicate, and object) must be resources uniquely identified by URI, except the object that can also be defined by a literal, which can be a string, number, or date.
    As shown in	Fig.~\ref{fig.rdf_triple_example}~(c), the object is a literal type, i.e., a string \textit{``Doe''}, which does not exist separately with this information.		
    RDF triples can be described using XML format, namely RDF/XML, as shown in~Fig.~\ref{fig.rdf_xml}.
    Other syntaxes, e.g., Terse RDF Triple Language (TURTLE) or Notation 3 (N3), can also be used for describing RDF triples~\cite{w3cRDF}.

 	\begin{figure}[t]
 		\centering
 		$\begin{array}{c}
 			\includegraphics[width=0.5\linewidth]{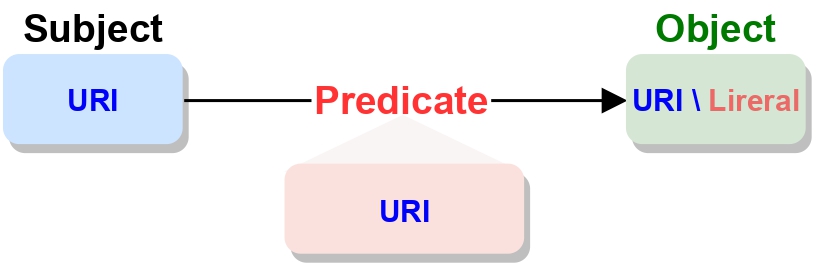}\\
            \parbox{0.8\linewidth}{\footnotesize (a) The RDF triple (statement): Subject-Predicate-Object~\cite{w3cRDF}}
			 			\vspace{8pt}\\
            \includegraphics[width=0.73\linewidth]{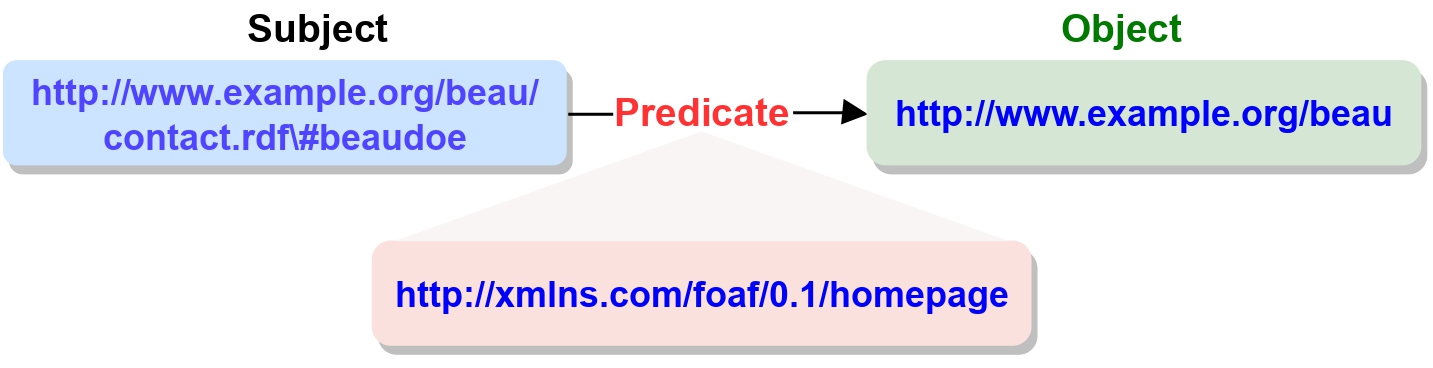}\\
 			\parbox{0.8\linewidth}{\footnotesize (b) The RDF triple of ``Beau Doe has a homepage http://www.example.com/beau''}
			 			\vspace{8pt}\\
 			\includegraphics[width=0.65\linewidth]{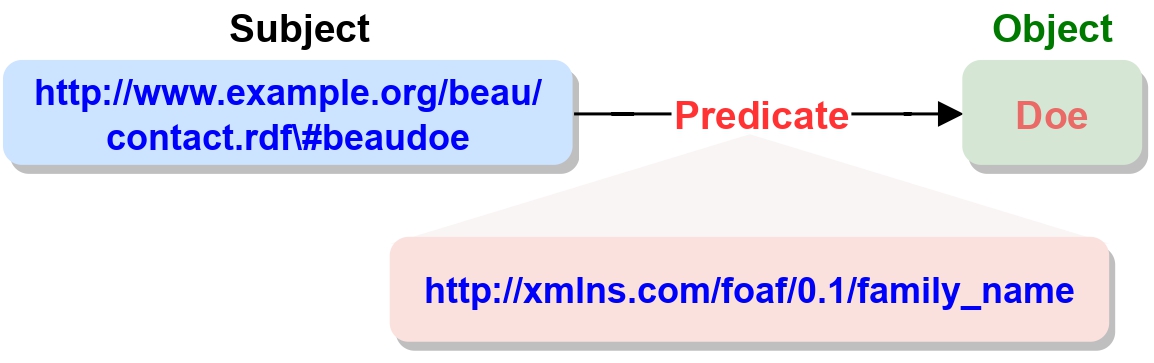}\\
 			\parbox{0.8\linewidth}{\footnotesize (c) The RDF triple of ``Beau Doe has family name Doe''}\\ 
 		\end{array}$
 		\caption{The RDF triple representation.}
 		\label{fig.rdf_triple_example}	
 	\end{figure}
 	
 	\begin{figure}[!t]
 		\centering
 		\includegraphics[width=0.95\linewidth]{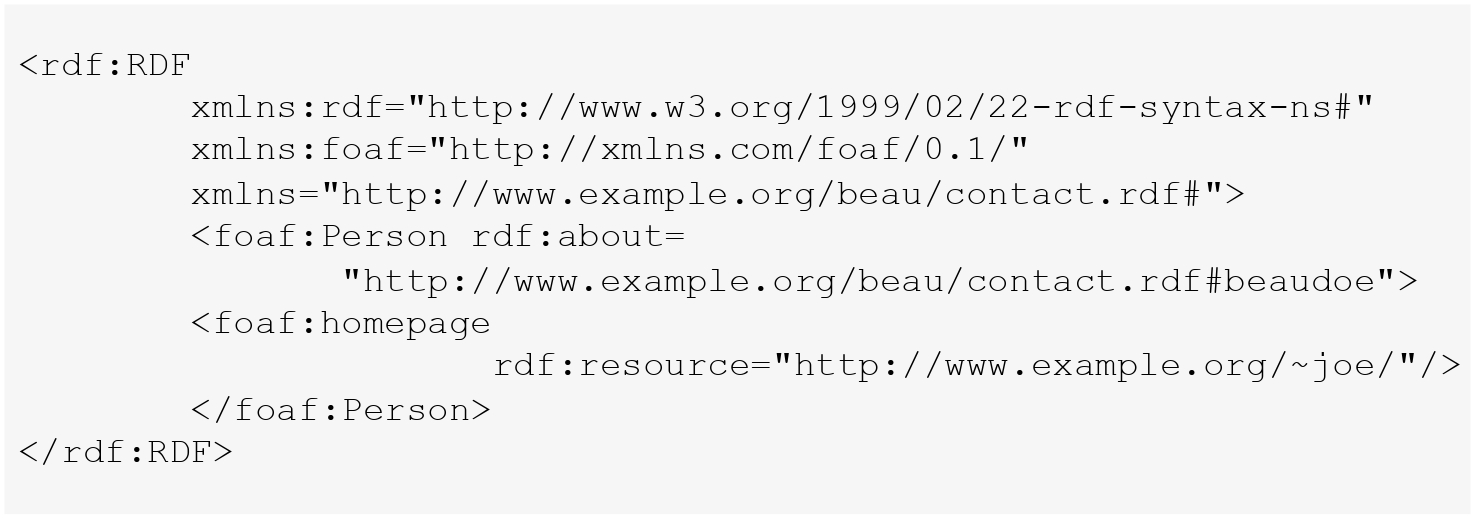}
 		\caption{The RDF/XML representation of ``Beau Doe has family name Doe''.}
 		\label{fig.rdf_xml}
 	\end{figure}
 	
 	\subsubsection{Taxonomies layer}
    This layer extends the RDF by defining how to model groups/classes of resources and the relations among these resources.
		In other words, the Taxonomy layer adds more vocabulary, i.e., definitions of concepts and relations used to represent knowledge,  to enable the classification of concepts~\cite{hitzler2009foundations}.
		Specifically, the technologies of this layer, e.g., the RDF Schema (RDFS), supply a vocabulary for describing classes and properties of RDF-based resources. 
		For that, the RDFS is also called RDF Vocabulary Description Language.		
		Let us return to the previous examples illustrated in Fig.~\ref{fig.rdf_triple_example}, Beau Does can belong to the \textit{person} class, the predicate can be the \textit{property} class, and both are sub-classes of the \textit{resource} class.
		In RDFS, the \textit{resource} class is the foundation of all other classes.		
 		
 	\subsubsection{Ontologies and rules layers} 	
 	The semantic web technologies aim to construct a global database, and thus this database may consist of partial or even contradicting information.
	 	For example, the word ``Jaguard'' can be a type of animal, but it can also be a car brand or a version of the Macintosh operation system, as shown in Fig.~\ref{fig.SWW_meaning}.
	 	To address this problem, the ontology layer aims to define a specific domain of discussion by extending the taxonomy layer's vocabulary. 
	 	As such, this layer takes responsibility for defining rules and knowledge about which relations make sense and are  allowed~\cite{hitzler2009foundations}.
	 	Here, an ontology language is the explicit representation of concepts and their interrelations~\cite{w3cRDFS}.
		For instance, the Web Ontology Language (OWL), which is based on description logic, can be used to define a  specification of a shared conceptualization, e.g., an abstract model of the domain, identified relevant concepts, and relations~\cite{gruber1993translation}.
		To support rules beyond the OWL and RDFS, the semantic technologies also consist of rule languages, which are used for expressing web rules that computers can execute.
		Currently, the Rule Interchange Format (RIF) and Semantic Web Rule Language (SWRL) are emerging standards at this layer.
 		
	\subsubsection{Querying layer} 
    The querying layer aims to provide an approach for obtaining information from RDF data, RDFS, and OWL ontologies.
		For that, W3C proposes the Simple Protocol and RDF Query Language (SPARQL), which is analogous to the Structured Query Language (SQL).
		The major difference is that while SQL uses string, e.g., the name of a table or column for querying data, SPARQL uses RDF triples and resources for this purpose. 
		In addition, SPARQL is also a protocol for accessing RDF-based data.
		
	\subsubsection{Unifying logic and other top layers}
	The unifying logic layer is essential to guarantee the reliability of data interchange between Query, Ontology, and Rule. 	
		The Cryptography Layer is a vertical layer offering cryptography techniques, e.g., digital signatures, to validate data sources, thus ensuring reliable inputs.
		To guarantee that the obtained knowledge can be trusted, the Proof and Trust layers are proposed to force all semantics and rules to be complied with before forwarding this information to the user interface and applications layer.
	
	To date, the semantic web technologies have been widely applied in practice, such as building Linked Data~\cite{lehmann2015dbpedia, sandhaus2010semantic, weaver2013facebook} and Knowledge Graph~\cite{noy2019industry}.
		However, the Web 3.0's main goal, i.e., providing a data framework for cooperating between machines and humans, is still not achieved due to immature technologies~\cite{hitzler2021review}.
		For example, the Unifying Logic, Proof, and Trust layers are not well-defined yet~\cite{w3cRDF}. 
		We will further discuss the state-of-the-art and challenges of the semantic web technologies for Web 3.0 in the following subsections.

\subsection{State-of-the-art in Semantic Web Technologies}
\subsubsection{Managing complexity and diversity}
Combining current web services to accomplish user seeks for specific functionalities while maintaining the highest possible quality of service (QoS) is involved with web service composition. Evolutionary computation (EC) methods are employed to address the complex computational nature of this problem. Moreover, for each individual combination demand, these methods can detect compound services that offer the quality of semantic matchmaking (QoSM), which is a near-optimal functional quality method, or non-functional quality ~\cite{wang2022using}. However, as the number of composition requests from users continues to rise rapidly, addressing one demand at a time is no longer efficient enough.  This way, the authors in~\cite{wang2022using} embarked on a groundbreaking study to address the challenge of managing diverse service composition demands as a collective problem ~\cite{wang2022using}.
Utilizing an estimation of distribution algorithm (EDA), they put forward a fresh approach based on a permutation-based multi-factorial evolutionary algorithm (PMFEA) called PMFEA-EDA, to solve this problem. Additionally, they introduced a technique for sharing knowledge across various service composition requests, along with the development of a new sampling mechanism. This mechanism of sampling aims to enhance the likelihood of service composition identification with superior quality in both single-tasking and multitasking scenarios.\par

\begin{figure*}[!]
    \centering
    \includegraphics[width=0.7\textwidth]{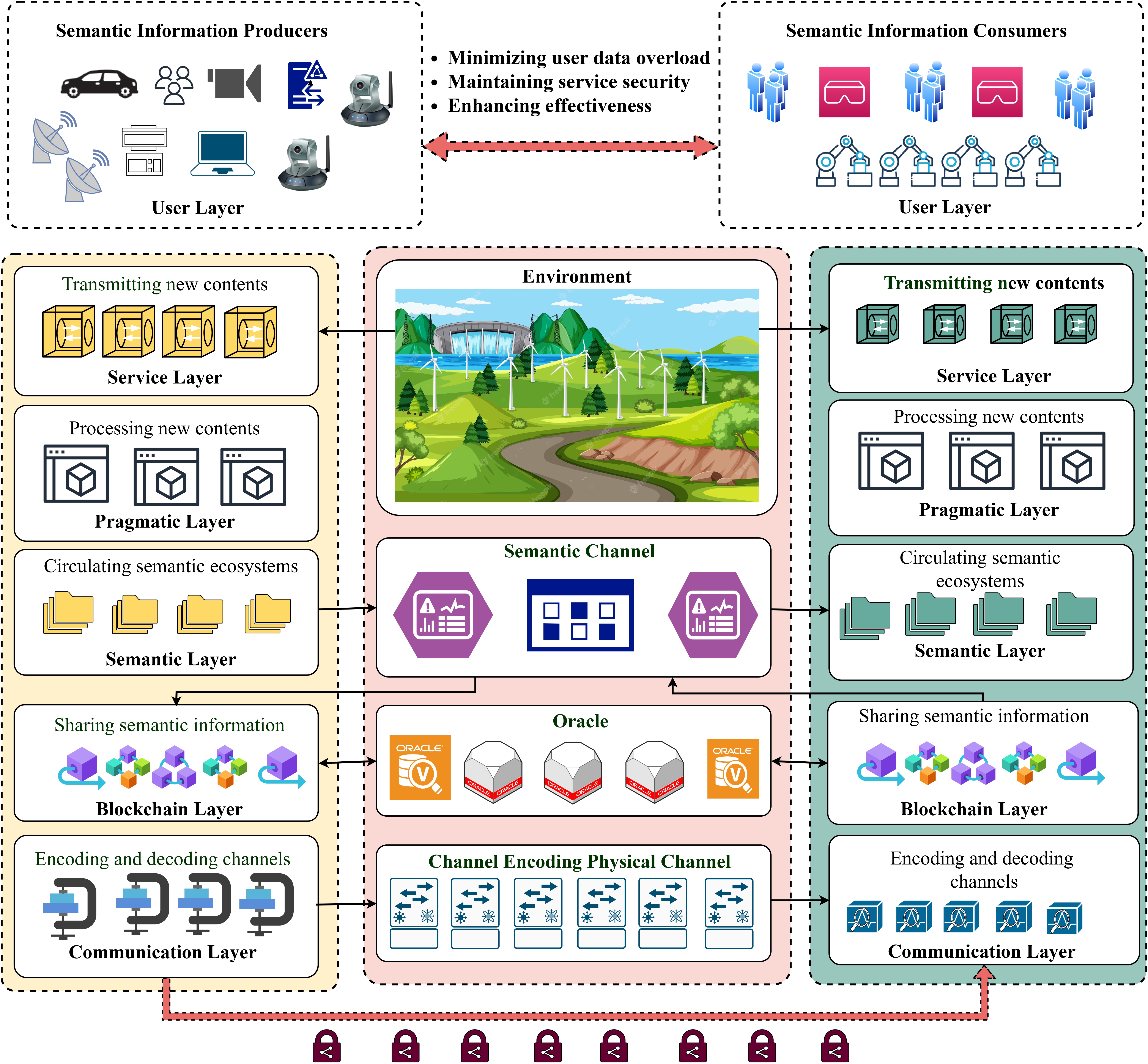}
    \caption{The framework proposed in~\cite{ lin2023unified} is based on the unified blockchain-semantic and enables Web 3.0.}
    \label{fig:SemBlock}
\end{figure*}

\subsubsection{Amalgamating content and structure}

The sparse and limited features found in web page summaries impact the web service's clustering performance. Moreover, the clustering process is further complicated by the influence of data sparsity and noisy features. To solve this problem, a clustering Web service called spectral clustering was proposed in~\cite{ kang2022web}. The authors combine the structured data and descriptive documents extracted from the service relationship network, this clustering web service can use unified content, structural information, and semantic representation (UCSI-SC) to improve the performance. Using Doc2vec and the service relationship network to learn structural-semantic information, this can extract content-based semantic details from descriptive documents. By pre-existing training of a service classification model, the method enables the amalgamation of content and structural-semantic information into cohesive features. Finally, the unified features are employed for spectral clustering to cluster Web services, achieving significant improvements in precision ~\cite{kang2022web}. \par

\subsubsection{Optimizing web service} 
The structural characteristics of the service network and the semantics of service content can be combined for large-scale service management. This way, to acquire initial features from service network relationships and service descriptions, two newly introduced models were employed.
Moreover, using a specific portion of the dataset including class labels, a combination of the MLP model and additional training extracted final features helpful for service clustering. Accordingly, using the last preprocessed features, the spectral clustering algorithm was then employed to achieve optimal service clustering through the dataset's characteristics. Another study introduced the DeepWSC framework, which enhances web service clustering by combining service composability and deep semantic features~\cite{zou2020deepwsc}. As conventional approaches to web service clustering have limitations, such as their inability to capture the semantic meaning of services and neglect of service composability, DeepWSC can overcome these limitations. Integrating deep learning techniques, it extracts nuanced semantic properties from service representations. Experimental evaluations demonstrate DeepWSC's superior performance, surpassing existing techniques in various validation indexes. Beyond its depiction, DeepWSC categorizes web services into clusters based on functional descriptions and relationships, fostering composability and functional diversity. This strategic approach aims to create a refined system, enabling seamless interactions and intricate functionalities through the thoughtful combination of diverse web services.

\begin{table*}
	\caption{Summary of the state-of-the-art in semantic web technologies for Web 3.0}
	\label{table_statesemantic}
	\begin{centering}
		\begin{tabular}{|>{\raggedright\arraybackslash}m{1.0cm}|>{\raggedright\arraybackslash}m{1.7cm}|>{\raggedright\arraybackslash}m{3.0cm}|> {\raggedright\arraybackslash}m{4.0cm}|>{\raggedright\arraybackslash}m{6.0cm}|}
			\hline 
			\multicolumn{1}{|>{\centering\arraybackslash}m{1.0cm}|}{\multirow{1}{*}{\textbf{Ref.}}} & 
            \multicolumn{1}{>{\centering\arraybackslash}m{1.5cm}|}{\multirow{1}{*}{\textbf{Application}}} &
            \multicolumn{1}{>{\centering\arraybackslash}m{2.9cm}|}{\multirow{1}{*}{\textbf{Research problem}}} &
            \multicolumn{1}{>{\centering\arraybackslash}m{2.9cm}|}{\multirow{1}{*}{\textbf{Solution}}} &
            \multicolumn{1}{>{\centering\arraybackslash}m{6.0cm}|}{\multirow{1}{*}{\textbf{Result}}}\\
			\hline 
			\hline

       \cite{ lin2023unified} & General web service & Secure data storage and efficient information interaction & Implementing a unified blockchain-semantic ecosystems framework with adaptive deep reinforcement learning-based sharding for resource optimization & Improving the interaction of information and achieving goal-oriented communication can enhance the efficiency, reduce latency, and facilitate the diversity of Web 3.0 services. \\ 	
        \cline{1-5}
                          
       \cite{ wang2022using} & General web services & Web service composition problem & Develop a new sampling strategy to enhance the likelihood of finding top-quality service compositions in both single-tasking and multitasking environments & The suggested approach surpasses two advanced single-tasking methods and one recent multi-tasking evolutionary computation-based technique in discovering superior solutions. \\ 	
        \cline{1-5}

       \cite{ kang2022web} & General web services & Functional affinity characterization challenges in complex web service networks and semantics & Utilize Doc2vec for content and network representation learning for structural semantics, enhancing Web service clustering & The proposed approach boosts precision by 4.78\% and recall by 5.4\% compared to the state-of-the-art method. \\ 	
        \cline{1-5}

       \cite{ zou2020deepwsc} & General web service & Clustering web services & Develop a heuristic framework that merges deep semantic features from service descriptions with composability features from invocation relationships, creating integrated features for web service clustering & DeepWSC excels in multiple metrics, surpassing state-of-the-art approaches for improved web service clustering on 8,459 real-world services. \\ 	
        \cline{1-5}

       \cite{ pang2022semantic} & Autonomous unmanned system & Heterogeneous cooperative autonomous unmanned systems & The framework combines ontology, deep learning, and SWRL rules for semantic-centered cloud control in autonomous unmanned systems & This work shows a feasible way to realize the cognitive ability of autonomous unmanned systems at task-level. \\ 	
        \cline{1-5}

		\hline                                               
	  	\end{tabular}
		\par\end{centering}
\end{table*}

\subsubsection{Efficient task cooperation and safety check} 
To enhance cooperative multi-unmanned ground vehicle (UGV) systems, a semantic-centered cloud control framework was introduced in ~\cite{pang2022semantic}. This framework involves the implementation of semantic modeling using the semantic network to establish an integrated design structure for tasks and environments. Additionally, with a combination of semantic web rule language (SWRL) and deep learning, a scene semantic information retrieval technique was employed to achieve task-level cloud task cooperation and scene understanding. One fundamental technology that plays a pivotal role in achieving the goals of Web 3.0 is blockchain. Blockchain facilitates transparent recording and decentralized content and ensures enhanced trust and security. Nevertheless, as the amount of on-chain recorded content expands rapidly and the user base continues to grow, the resource consumption associated with computing and storage escalates. This poses affordability challenges for users due to increased operational costs. A promising approach is to focus on the analysis of semantic information in content, allowing for the precise conveyance of desired meanings while minimizing resource consumption. Fig.~\ref{fig:SemBlock} demonstrates a comprehensive framework that was proposed to enable wireless edge intelligence in the context of Web 3.0~\cite{ lin2023unified}. As is observed in Fig.~\ref{fig:SemBlock}, the platform comprises six important elements aimed at facilitating the exchange of semantic demands. To implement interactions within Web 3.0 ecosystems while ensuring both off-chain and on-chain service security, a proof of semantic mechanism based on Oracle technology was introduced. Additionally, an adaptive sharding mechanism based on deep reinforcement learning, implemented on the Oracle, was designed to enhance interaction efficiency and enable Web 3.0 frameworks to effectively address diverse semantic requests~\cite{ lin2023unified}. \par

 Another study presented a semantic method to combine diverse data within a building information modeling (BIM) ecosystem. This approach aims to enhance safety checking automatically through the utilization of SPARQL-based reasoning techniques~\cite{li2022bim}. The approach involves converting them into integrated ontology instances, extracting threats from BIM and sensor information, and developing interconnected ontologies for subway construction safety checking. In this way, the automation of converting text-based rules within checking rules empowered by SPARQL-based can be achieved. As a result, the proposed platform can improve automated safety checking and knowledge sharing.\par

 \textit{Summary}: Semantic web technologies play a vital role in Web 3.0 for a range of capabilities. These technologies can facilitate the seamless integration of web services, ensuring high-quality service composition. They can cluster and manage web services effectively, improving discoverability and utilization within Web 3.0 ecosystems. Semantic modeling and scene understanding can enhance cooperative systems like multi-unmanned ground vehicle (UGV) systems for intelligent decision-making. Resource consumption challenges in Web 3.0 can be addressed through semantic information analysis, reducing resource usage while ensuring secure interactions. Additionally, heterogeneous data integration and automated safety checking can be achieved using semantic web technologies, improving knowledge sharing and real-time monitoring. Table~\ref{table_statesemantic} summarizes the current state-of-the-art semantic web technologies tailored for Web 3.0. It provides a condensed overview of the advancements, methodologies, or key findings in the field of semantic web technologies relevant to the context of Web 3.0.\par

\section{3D Interactive Web Technology}
\label{sec:3D_web}

\begin{figure*}[!]
    \centering
    \includegraphics[width=0.8\textwidth]{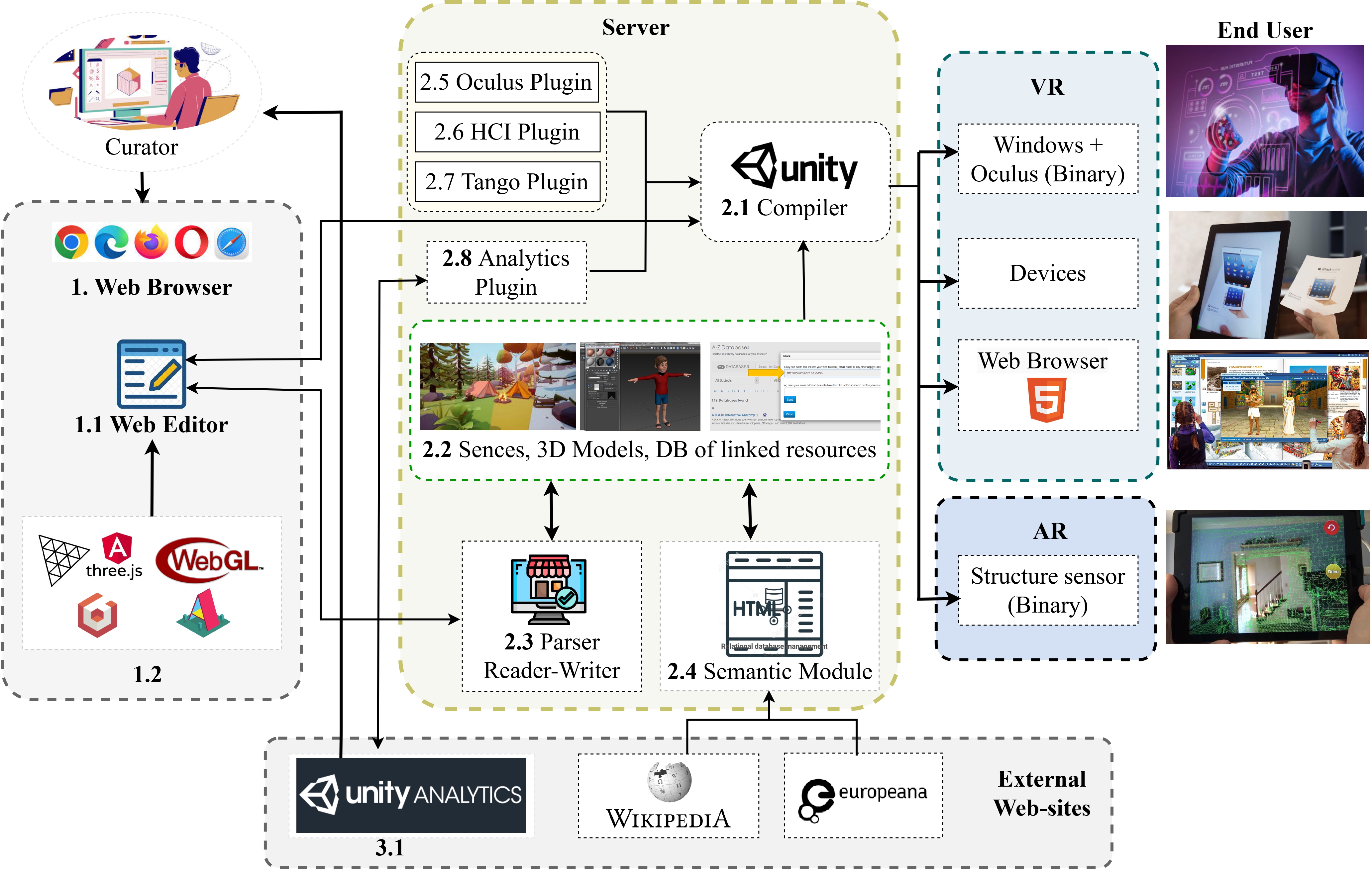}
    \caption{A high-level design for a server-based 3D interactive web platform for 3D cultural worlds~\cite{digiart}. The design is intended to facilitate the building of new 3D cultural worlds and comprises a web editor, gaming assets, and semantics. It has been created to make it easier for curators and non-programmers to develop interactive experiences and exhibitions without requiring extensive programming knowledge.}
    \label{fig:3d}
\end{figure*}

\subsection{Background of 3D Interactive Web Technology}

The emergence of 3D interactive web technology, sometimes known as ``Web3D'' has brought in a new era of engaging and dynamic online experiences for users. It enables web users to engage with immersive and dynamic 3D content. Furthermore, it creates incredible new opportunities for online creativity, communication, and engagement through the use of 3D images, by real-time interactivity~\cite{patterson2016interactive}. The conventional 2D web has limited contents, such as texts, photos, and videos, Therefore, the 3D interactive web offers direct access to immersive worlds through web browsers by integrating Hypertext Markup Language version 5 (HTML5) www.html.com, Web Extended Reality (WebXR) (www.immersiveweb.dev), Three.js (www.threejs.org), Babylon.js (www.babylonjs.com), PlayCanvas (www.playcanvas.com), and Extended 3D Technology (X3D) (www.web3d.org). These cutting-edge technologies enable the 3D interactive web to be easily accessible on a wide range of platforms and devices. Fig.~\ref{fig:3d} illustrates a high-level design for a 3D interactive web platform. The design includes a web editor, game assets, and semantics to assist in the development of innovative 3D cultural worlds. This design thus simplifies the development of interactive experiences and exhibitions by curators and non-programmers who lack programming expertise.

Web graphics library (WebGL) enables the representation of interactive 3D graphics, visualizations, games, and virtual reality experiences directly in web browsers, without any plugins. It is crucial for building engaging virtual worlds where users can participate in 3D online experiences. The primary purpose of WebGL is to connect JavaScript and the computer's GPU, allowing for rapid real-time processing of complex 3D scenarios~\cite{baruah2021ar}. It uses custom shader programs written in OpenGL Shading Language (GLSL) to control how 3D objects look at the final output. It also provides users to navigate and interact with a virtual world to explore objects and environments in real-time with 3D materials,  that were previously restricted to the traditional web. 

Fig. \ref{fig:WebGL technology} illustrates the WebGL technology utilized in the 3D web platform. In particular, the figure consists of three layers, each layer playing a crucial role in enabling the creation and rendering of 3D graphics in web applications. On the top layer, middleware layer plays a pivotal role by providing developers with essential tools and libraries like three.js. This layer acts as an interface, allowing developers to focus on creating 3D experiences. Moving to the middle layer,  WebGL API provides a rich JavaScript middleware ecosystem which is used to bridge the gap between the middleware and the rendering engine to ensure an enjoyable browsing experience~\cite{fan2023large}. Finally, at the bottom layer, WebGL utilizes OpenGL, for translating high-level requests from the WebGL API into hardware-specific commands, ensuring 3D content across a variety of devices and systems seamlessly~\cite{stemkoski2021developing}. The next sub-section delves into an exploration of the roles of 3D interactive web technology, and how this innovative technology is revolutionizing user experiences in Web 3.0.
\begin{figure}[htb]
    \centering
   \includegraphics[width=1.0\linewidth]{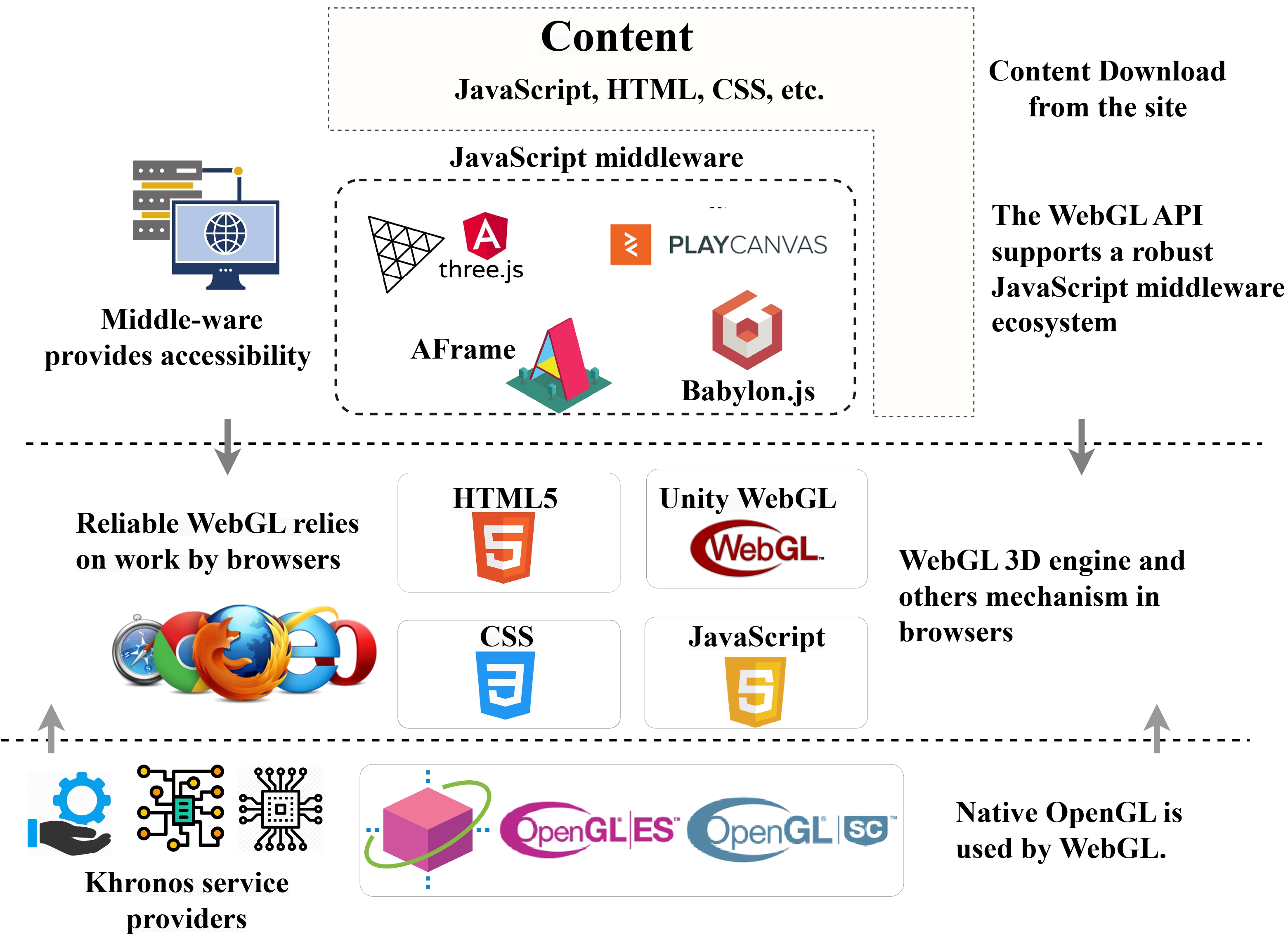}
    \caption{WebGL technology is used in 3D web platforms~\cite{varvara}. On the top layer, middleware provides accessibility for the developers (e.g., three.js library), while WebGL API provides a powerful foundation for a rich JavaScript middleware ecosystem. At the second layer, the browser provides a WebGL 3D engine with other mechanisms. At the bottom layer, WebGl uses native services (i.e., OpenGL) a cross-platform for producing 3D.}
    \label{fig:WebGL technology}
\end{figure}

\subsection{The Role of 3D Interactive Web Technology in Web 3.0}
The 3D interactive web is an essential component of the implementation of Web 3.0, creating its environment while rendering it a reality. It has a huge influence on the web's growth toward Web 3.0, which is designed to be more immersive, intelligent, and decentralized~\cite{Zdrzalek2021}. Web 3.0 users are expected to experience dynamic and exciting information which is efficient and can be achieved using 3D interactive web. Users can immerse themselves in virtual environments where they interact with virtual items and explore in real-time by allowing 3D experiences in web browsers~\cite{discher2019concepts}. This thus can fill the gap between the virtual and real worlds, giving users personalized and exciting experiences in the Web 3.0 era. In short, the roles of 3D interactive web technologies in Web 3.0 are:

\begin{itemize}
    \item Enhance web users' experiences by using 3D visuals, VR/AR headsets, haptic devices, and recognition sensors~\cite{kelven}. These technologies have become essential for developing interactive worlds with an atmosphere of reality and immersion. This technology improves the web experience for all users by making it more interesting and realistic.
   
    \item Enable the creation of new applications and services that make effective use of the web's spatial and semantic features. These applications include virtual museums, gaming platforms, e-commerce solutions, real estate interfaces, tourism platforms, and more~\cite{pixelplex}. Leveraging these semantic features enables the development of dynamic and engaging experiences, providing users with a diversified and immersive online environment across several domains and industries.

    \item Boost user and community involvement across a variety of platforms and devices. Shared experiences and seamless engagement can be achieved using 3D avatars, social networks, live streaming, and related technology~\cite{deloitte}. These technologies are essential to building a digital world where individuals can communicate, work together, and develop connections anywhere within any device. Incorporating such components makes online engagement more dynamic and exciting, encouraging an atmosphere of community and shared engagement.
    
\end{itemize}

 \begin{figure*}[!]
    \centering
    \includegraphics[width=0.8\textwidth]{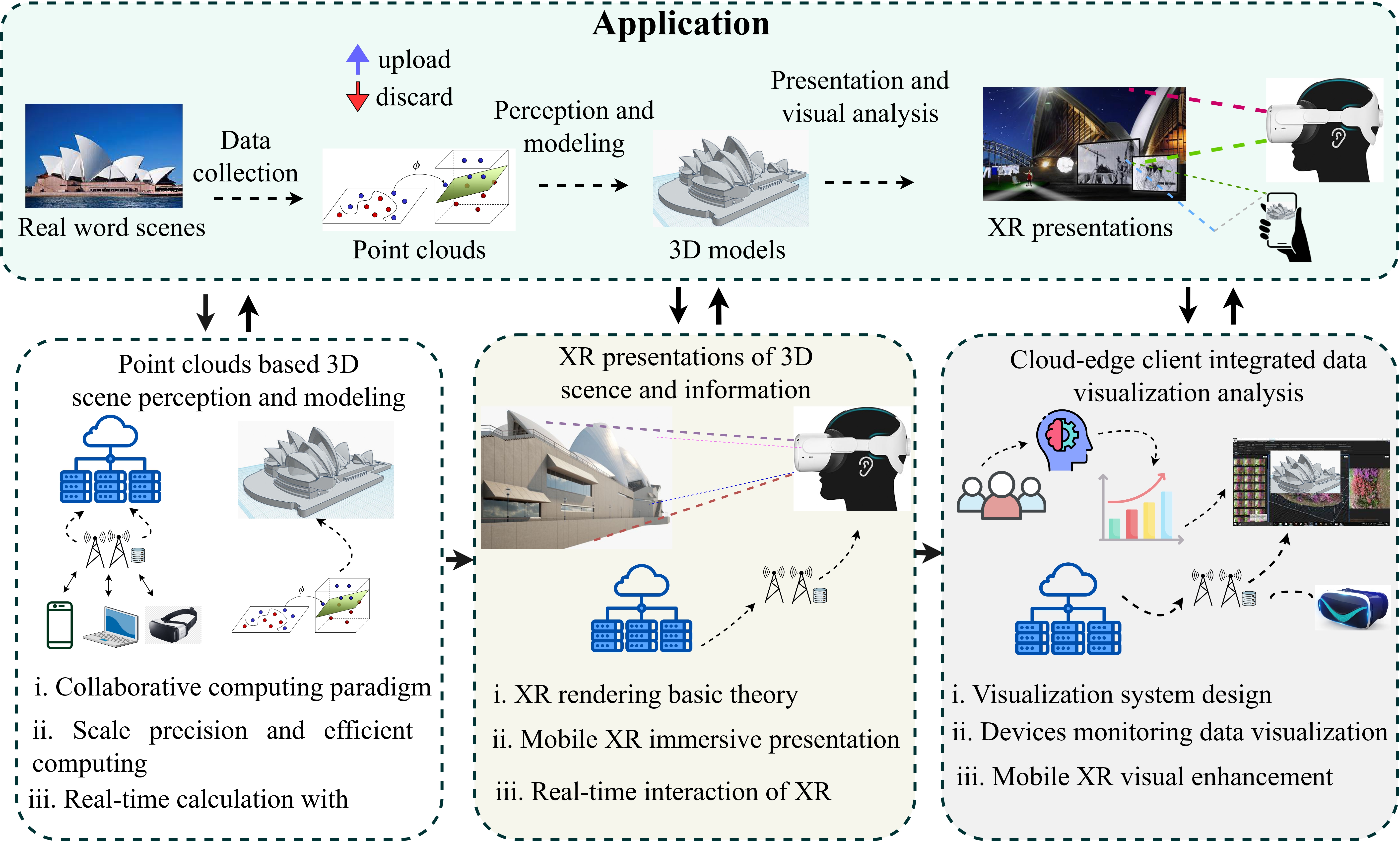}
    \caption{The proposed 3D scene modeling and XR presentation of the cloud-edge-client architecture~\cite{wu2023perspectives}. The process involves using point cloud-based 3D scene perception and modeling, collaborative computing paradigm, and XR presentation. The different stages of the process work together to optimize task allocation, ensure system openness, scalability, and controllability, and present the 3D digital twin model in an immersive and interactive way.}
    \label{fig:new fig}
\end{figure*}

 \subsection{Recent Development on 3D Interactive Web Technology}

\subsubsection{Extended 3D technology (X3D) for 3D interactive web}

The development of the 3D web encouraged an evolution in the realm of virtual reality and revolutionized the way users use the web. Users can immerse themselves in a full 3D revolutionary reality using cutting-edge technology like VR and AR~\cite{van2022reconstruction}. In contrast to the conventional online, the users encounter challenges such as interoperability, limited interaction, and scalability~\cite{flotynski2019semantic}. However, these challenges can be mitigated using ontology-based 3D semantic knowledge representations which enable interoperability, improve 3D models, enhance user interaction, and handle 3D data efficiently~\cite{qin2016ontology}.

Recently, the authors in~\cite{flotynski2019semantic} proposed an X3D ontology technique for 3D interactive web. The considered framework combines ontologies and semantic technologies with 3D modeling. In particular, the 3D content ontology contains information about specific 3D models and scenes, including dataset types, encoding standards, and activities.  The framework focuses on creating ontologies to capture semantic information for enriching 3D models. It also incorporates mechanisms for storing, querying, and semantic 3D data, enabling advanced search semantic capabilities,  interoperability, and data sources. The authors highlighted that the proposed framework presents a new opportunity for the future of semantic 3D content.

A similar X3D data model was proposed in~\cite{kim2020representation} for smart cities. In the considered framework, the XML model is used to develop a 3D virtual atmosphere for healthcare applications. In particular, X3D and Hanim are used to represent the 3D virtual world and the 3D avatar. The 3D environment is interconnected with real-time GPS data, and the avatar is monitored based on its GPS location. The data model is then defined with XML and handled with an X3D browser. The authors showed that the proposed framework can be used to observe the user's health data in real life.

\subsubsection{The 3D model for 3D interactive web}

The use of 3D models in 3D interactive web brings numerous advantages in generating realistic visuals in AR and VR applications. However, the advancement of AR and VR technologies has brought challenges for existing WebGL 3D JavaScript libraries, such as data transmission, limited mobile browser computation, and latency in handling 3D models~\cite{ma2021research}. Leveraging advancements in rendering 3D technologies and optimizations can help reduce latency and ensure the smooth rendering of 3D models and animations. As a result, the authors in~\cite{li2020rendering} proposed a separate animation data for Web3D. In the considered framework, the data model is divided into animation and topological sequences enabling the efficient loading of essential data during the initial rendering. The mesh model is exported in JD file format, and data is stored separately based on animation tracks. A multi-granular model support is employed for continuous animation support. The Three.js techniques are optimized to accomplish asynchronous uploading data and on-demand rendering that significantly reduces the initial loading time and improves efficiency, especially in interaction scenarios with multiple animations. The authors demonstrated that the proposed optimization method reduces the latency in model data transmission and improves complex interactive scenarios.

Another 3D model was proposed in~\cite{fan2021interactive} for urban planning. In the considered framework, the authors proposed a dynamic platform for 3D building which is essentially made up of user interaction, automation, and real-time modeling. The 3D modeling is efficiently performed by an object detection network. The Convolutional Neural Network (CNN) was implemented to develop the framework. The authors evaluated that the proposed framework can offer the 3D modeling community with cost-effectiveness, interaction, and efficiency as a possible solution.

\subsubsection{HTML5 for 3D interactive web}

HTML5 is another core technology that plays a vital role in 3D interactive web. It enables developers to build spectacular 3D interactive experiences directly in the browser with the advent of WebGL (i.e., a JavaScript API for displaying 3D visuals) in the browser. HTML5 can address challenges in 3D online learning by offering individuals a dynamic and engaging experience. Due to security issues and the demand for multi-device user experience, HTML5 replaced Adobe Flash for online learning platforms~\cite{tabares2021html5}. Therefore, the authors proposed an HTML5-based 3D animation framework for Web3D in \cite{lei2020unified}. The authors explored a human-centered online system to control engineering experimentation. In the considered framework, HTML5 is used for controlling the immersive and interactive learning environment to ensure security and designing the graphical user interface. In particular, by replacing the outdated flash plug-in with HTML5, compatibility, and accessibility are greatly enhanced, creating a way for learners to access the system comfortably. The authors highlighted that the proposed framework enhances the seamless integration of 3D features, real-time interaction, and online learning experiences/opportunities in 3D interactive web.

\subsubsection{WebXR for 3D interactive web}
WebXR is one of the important elements in 3D interactive web, which provides a standardized API for immersive experiences that are easily accessible through web browsers. It has transformed how we experience and engage with 3D information on the web, linked with many devices and platforms. WebXR can help tackle problems such as performance, accessibility, and compatibility in the 3D web by offering a standard approach to developing immersive experiences that function across many devices and platforms~\cite{maclntyre2018thoughts}. Therefore the authors in~\cite{hanfati2022design} proposed a WebXR-based health application framework in Web3D. The considered framework, developed on different devices, (i.e., phones and tablets), gives an immersive experience. In particular, the authors used WebGL and WebXR JavaScript libraries, for interactive 3D content without using any web UI/UX components. The WebGL engine is used to develop 3D medical object assets, scenarios, and extended reality components. Users can interact with the 3D animated objects using devices to ensure an immersive experience. All medical scenarios are stored in a database using WebXR dynamically on the user's device. Users can access the environment individualized to their device for an enhanced user experience. The authors showed that the proposed method potentially incorporates interactive elements for an engaging educational experience with accessibility, and cost-effectiveness.

\begin{table*}
	\caption{Summary of 3D interactive approaches for Web 3.0}
	\label{table_3Dapproach}
	\begin{centering}
		\begin{tabular}{|>{\raggedright\arraybackslash}m{1.0cm}|>{\raggedright\arraybackslash}m{2.0cm}|>{\raggedright\arraybackslash}m{3.7cm}|> {\raggedright\arraybackslash}m{4.0cm}|>{\raggedright\arraybackslash}m{5.0cm}|}
			\hline 
			\multicolumn{1}{|>{\centering\arraybackslash}m{1.0cm}|}{\multirow{1}{*}{\textbf{Ref.}}} & 
            \multicolumn{1}{>{\centering\arraybackslash}m{1.5cm}|}{\multirow{1}{*}{\textbf{Application}}} &
            \multicolumn{1}{>{\centering\arraybackslash}m{2.9cm}|}{\multirow{1}{*}{\textbf{Research problem}}} &
            \multicolumn{1}{>{\centering\arraybackslash}m{2.9cm}|}{\multirow{1}{*}{\textbf{Solution}}} &
            \multicolumn{1}{>{\centering\arraybackslash}m{4.0cm}|}{\multirow{1}{*}{\textbf{Result}}}\\
			\hline 
			\hline 
   
             \cite{flotynski2019semantic}  &  Education & Addressing 3D technology integration in immersive virtual and augmented reality scenarios  & Semantic ontology for 3D representation & Enhance 3D representation on the Semantic web \\ \cline{1-5}

             \cite{kim2020representation}    & Healthcare  & Addressing the real-time data monitoring in smart city  &  X3D visualization engagement for 3D health management & Significantly managed and controlled health information in a virtual environment \\
			\cline{1-5}

            \cite{li2020rendering} & Animation data & Addressing data transmission limitations, poor computing, and delay in a 3D model & Large scale 3D models to improve the data request and loading process & Significantly reduced the latency,  data transmission, and rendering up to 24.72\% \\
         	\cline{1-5}
   
			\cite{fan2021interactive} &  Design and architecture & Eliminating time and labor expenses on a 3D interactive platform & 3D model for designing building and CNN for object detection & Significantly enhanced user interaction and usability in 3D modeling \\
			\cline{1-5}

           \cite{lei2020unified} &   Education & Addressing remote virtual learning on the 3D web & 3D animation for remote virtual experiments and HTML5 for graphical user interface & Enhanced online learning experiences and expanded 3D opportunities for remote virtual environment \\
			\cline{1-5}

           \cite{wu2023perspectives} & E-governance & Addressing computational power on large-scale 3D scene modeling & Large-scale 3D models presentation, and visualization using WebXR & Significantly reduced large-scale 3D scene high bandwidth and latency issues  \\ 
           \cline{1-5}
           
			\cite{hanfati2022design} &  Health informatics & Addressing remote medical learning practice & Virtual medical education and interaction using WebXR representation & Increase user engagement up to 4.64\% (by survey) satisfaction in medical education and training on the multi-cross XR platform\\
			\cline{1-5}

           \cite{guo2022using} &   E-learning & Addressing engagement and efficiency issues on 3D interactive web  & Game-Based WebXR representation using VR and AR & Experiment was conducted with 17 participants to evaluate the engagement of students in primary school learning using 3D models and audio recordings \\
			\cline{1-5}

   		\cite{gonccalves2022gait} &  Health informatics & Considering Gait disorders quality and efficiency & Unity 3D for interaction between the user and the VR scenario and Azure Kinect for body tracking & Developed an app for physiotherapists to track users' rehabilitation with resulting highest value up to 77.5\%  \\
			\cline{1-5}
   
			\hline                                                       
	  	\end{tabular}
		\par\end{centering}
\end{table*}

Another WebXR-based framework was proposed in~\cite{guo2022using} for learning in Web3D. In the considered framework, the system's objectives are achieved by constructing functional stacks using three layers. In particular, the learning system uses 3D interactivity on a screen, with movable 3D models and audio recordings, to enhance accessibility. The mobile VR interaction enables users to discover a 3D space by moving their mobile devices. The framework incorporates AR interaction through Simultaneous Localization and Mapping (SLAM), allowing learners to load a 3D object view and utilize their mobile device camera to scan a surface for AR. For rendering XR material, A-frame VR and Google Model-viewer have been employed. The framework uses a client-server model, with server-side and reusable client-side functions. The system runs on a Windows server and includes two databases for user interactions. The authors demonstrated that the proposed framework can improve the learning system in the Web3D area. Moreover, the authors in~\cite{wu2023perspectives} proposed a similar client-based cloud-edge architecture for Web3D. The authors concluded that their proposed framework will address large-scale 3D scene challenges. Fig.~\ref{fig:new fig} illustrates the proposed cloud-edge-client framework. The framework explores the challenges and opportunities of using point cloud data to create 3D digital twins of real-world scenes and present them in immersive and interactive ways using XR technologies. Furthermore, it also addresses inefficiencies in real-world industrial use and improves perception of scenes and comprehension, 3D model generation, and large data analysis and presentation intelligence.

\subsubsection{Unity 3D for future 3D interactive gaming}
Unity 3D is another important technology in 3D applications because it provides an efficient and versatile platform for generating immersive visual online experiences. Its extensive number of tools and capabilities empowers developers to bring their creative ideas into action and provide compelling 3D material to consumers. One of the key challenges it can solve is cross-platform compatibility, allowing developers to create 3D programs that can operate on a number of devices and operating systems. Therefore, the authors in~\cite{gonccalves2022gait} proposed a 3D application for health monitoring and implemented a web-based framework for gait rehabilitation and training exercises using Azure Kinect and 3D features. The Azure Kinect camera captured images that were analyzed by Microsoft's embedded AI algorithm to track human body joints.  3D interface utilized a body tracking software development kit (SDK) to replicate patient movements to an avatar. The proposed framework enables patients to access information, visualize training data, and make decisions regarding health conditions. Patients accessed and completed training exercises through the 3D platform. The authors assessed the effectiveness of the framework and encouraged users to follow physical rehabilitation goals using the use of VR technology. 

 \textit{Summary}: This section focuses on the application of 3D interactive technology in the development of Web 3.0 in the future. This technology can dynamically improve the seamless integration of virtual reality experiences and enable high-quality immersion in virtual worlds. Despite of its advantages, it also presents challenges for developers, such as producing realistic and engaging 3D displays for users and developing reliable and efficient algorithms for processing, presenting, and transmitting 3D data over the Internet. All these challenges will be carefully considered when integrating enabling technologies (i.e. WebGL, HTML5, WebXR, Three.js, Babylon.js, and X3D) in Web 3.0 in the future. A summary of the 3D interactive approaches for Web 3.0 is presented in Table~\ref{table_3Dapproach}.

\section{Open Issues and Emerging Technologies}
\label{sec:Dis}
In this section, we discuss the inherent challenges of incorporating emerging technologies to improve the performance of Web 3.0. We also introduce ideas to address the complexities associated with these novel fusions. Our main focus centers on challenges arising from the integration of one or several technologies, such as blockchain, semantic web, 3D interactive, and IoT, as the infrastructure for Web 3.0.

\subsection{Open Issues}
\subsubsection{Security and privacy}
Security and privacy are paramount in the context of Web 3.0 due to its transformative nature as an advanced, decentralized, and virtual internet \cite{winter2021s}. The integration of technologies such as blockchain, IoT, AI, semantic web, and interactive 3D Internet results in a highly interconnected and dynamic network space. Within this intricate environment, various actors with distinct objectives, strengths, and weaknesses contribute to a complex ecosystem. Ensuring the safety, privacy, and confidence in data and transactions is crucial for the progress of Web 3.0, safeguarding the rights and benefits of users, designers, and suppliers alike. The emphasis on accuracy, integrity, and confidentiality in Web 3.0 applications enhances overall quality, fostering dependability and credibility. Additionally, prioritizing the protection of confidentiality and privacy promotes collaborative efforts among stakeholders, fostering synergies. The reliability of data and transactions is pivotal for the creative and innovative development of Web 3.0 applications, ultimately enabling scalability, efficiency, and minimizing associated costs and risks.

\subsubsection{Interoperability, scalability, and efficiency}

 Interoperability, scalability, and efficiency are crucial for the success and widespread adoption of Web 3.0, as they determine the practicality and effectiveness of this new Internet era \cite{hatzivasilis2018interoperability}. Interoperability ensures that different blockchain networks and DApps can seamlessly communicate and transact with one another, fostering a cohesive ecosystem rather than isolated platforms. Scalability is essential to accommodate the growing number of users and transactions on the network without compromising performance, which is vital for real-time applications like streaming and gaming in the Web 3.0 space. Efficiency in processing transactions and using resources is key to making Web 3.0 both environmentally sustainable and cost-effective for users. However, achieving these objectives in a decentralized environment presents significant challenges. Ensuring interoperability across diverse and often incompatible blockchain protocols is a complex task, requiring standardized frameworks and technologies. Scalability poses difficulties as increasing the capacity of decentralized networks often leads to challenges in maintaining security and decentralization. Furthermore, enhancing efficiency, especially in consensus mechanisms like Proof-of-Work, which are energy-intensive, is crucial to address environmental concerns and improve transaction speeds. Thus, while interoperability, scalability, and efficiency are foundational for the functionality and growth of Web 3.0, they also represent some of the most significant technical hurdles that need to be overcome.

\subsubsection{Energy efficiency and performance optimization}

To fully harness the capabilities of Web 3.0, it is crucial to establish a balanced synergy between the energy consumption and operational efficiency of the blockchain and IoT devices that are pivotal components within the computational and communicative aspects of Web 3.0. This equilibrium presents a formidable challenge for several compelling reasons. Firstly, blockchain technology is well-known for its inherent energy-intensive characteristics. This is predominantly attributed to the multitude of nodes required for transaction validation and ledger maintenance, each necessitating substantial computational resources, consequently leading to substantial energy consumption. Secondly, it is important to consider that IoT devices often operate on limited battery power, necessitating prudent management of energy resources to ensure sustained functionality. Integrating these IoT devices into blockchain networks could potentially deplete their power reserves, thus reducing their operational lifespan. Thirdly, Web 3.0 is poised to encompass a multitude of diverse applications and services, each entailing specific requirements for the involved blockchain and IoT devices. 
For example, certain applications within the Web 3.0 ecosystem may demand high-performance capabilities, while others may prioritize energy efficiency. A skewed trade-off between energy consumption and performance could potentially lead to detrimental consequences for the advancement of Web 3.0. Excessive energy consumption within blockchain networks may discourage user engagement and jeopardize the sustainability of Web 3.0. Additionally, if IoT devices cannot effectively conserve energy, their ability to participate in Web 3.0 applications and services could be restricted.

\subsection{Emerging Technologies}
\label{sec:Fut}

The integration of cutting-edge technologies assumes a pivotal role in shaping the trajectory of Web 3.0's evolution. This section delves into a spectrum of emerging technologies poised to significantly advance the development of Web 3.0. 


\subsubsection{ WebAssembly (Wasm)}

Web 3.0 is envisaged as a transformative force in shaping a more intelligent, decentralized, and immersive online environment, as noted in a work on blockchain~\cite{ragnedda2019blockchain}. This evolution necessitates web applications and services to augment their efficiency, resilience, and adaptability across diverse platforms and devices. An instrumental solution to address these challenges is WebAssembly~\cite{webassembly_tutorial}, a binary instruction format for a virtual computer that can operate within web browsers. WebAssembly liberates online programs and services from the limitations imposed by intermediary layers of abstraction between JavaScript and the CPU, allowing them to fully harness the computational capabilities of the CPU~\cite{webassembly_tutorial}. The implications of WebAssembly for the online platform are extensive, as it empowers web-based client applications that previously struggled to attain such performance to execute code written in multiple languages~\cite{mdn_webassembly}. Importantly, WebAssembly is engineered to seamlessly coexist with JavaScript, facilitating the incorporation of WebAssembly modules into JavaScript applications through the use of WebAssembly JavaScript APIs. This integration enables users to leverage WebAssembly's capabilities while harnessing the versatility and adaptability of JavaScript within the same applications, even for those who may not be well-versed in writing WebAssembly code~\cite{mdn_webassembly}.

\subsubsection{ Quantum computing}
In the pursuit of establishing a decentralized digital economy, Web 3.0 stands as a pivotal element in the ongoing process of digital transformation. Constructed upon computing power networks, distributed data storage, and blockchain technology, Web 3.0 is evolving concurrently with the swift implementation of quantum devices \cite{xu2023quantum}. This breakthrough stems from the use of qubits, quantum counterparts to classical bits, which can exist in multiple states simultaneously. This unique property enables quantum computers to tackle problems that would be impossible for classical systems, opening up a vast realm of possibilities for the future of Web 3.0. In the realm of Web 3.0, quantum computing holds immense potential to enhance security and data integrity. By harnessing its power to create quantum-resistant encryption methods, we can safeguard sensitive information in the decentralized Web 3.0 landscape. Additionally, quantum computing could empower decentralized platforms to handle complex algorithms and massive datasets with greater efficiency, revolutionizing fields like decentralized finance (DeFi) and supply chain management. However, the advent of quantum computing also presents formidable challenges, particularly in the area of cybersecurity. Its ability to break current cryptographic protocols necessitates the development of new security paradigms to protect data and transactions in the decentralized web. This intersection of quantum computing and Web 3.0 will be a crucial area for innovation, driving the creation of robust security measures and potentially redefining how data is processed and secured in the evolving digital landscape.

\subsubsection{Zero-knowledge proofs (ZKP)}
Zero-knowledge proofs (ZKPs) are a revolutionary cryptographic method enabling one party to prove to another that a statement is true without revealing any information beyond the validity of the statement itself~\cite{sun2021survey}. This technology is particularly significant in the context of privacy and security in digital transactions. In Web 3.0, characterized by its decentralized nature and heightened focus on user privacy and data sovereignty, ZKPs offer immense potential. They enable the verification of transactions or data authenticity within blockchain networks without exposing the underlying data, thus maintaining privacy and confidentiality. This is especially beneficial for decentralized finance (DeFi), secure voting systems, and identity verification processes, where the need to validate information without compromising privacy is paramount. Moreover, ZKPs can enhance the scalability of blockchain networks by enabling more efficient data processing and reducing the amount of information that needs to be stored on-chain. The ability to prove the correctness of transactions without revealing their contents could greatly reduce the computational load and storage requirements, addressing some of the scalability challenges faced by current blockchain technologies. As Web 3.0 continues to evolve, integrating ZKPs into its framework could significantly advance its capabilities in terms of privacy, security, and efficiency, making it a key technology for the future development of the decentralized web.

\subsubsection{Generative AI}
Generative AI, a facet of artificial intelligence capable of generating novel content spanning text, visuals, and music, assumes substantial importance within the evolving sphere of Web 3.0. This significance rests on a multitude of compelling rationales~\cite{shen2023artificial}. Firstly, it acts as a catalyst for the development of inventive and groundbreaking Web 3.0 applications. For instance, generative AI finds utility in crafting applications that produce personalized content, aiding users in the creation of their digital assets, and facilitating novel, immersive interactions within the digital domain. Secondly, generative AI contributes to the augmentation of Web 3.0 applications' performance and scalability. It plays a pivotal role in conceiving innovative data compression algorithms, streamlining the performance of Web 3.0 networks, and introducing advanced security features to fortify Web 3.0 applications. Thirdly, generative AI endeavors to simplify and enhance the user experience in Web 3.0, manifesting through the creation of intuitive tools and interfaces that streamline user interactions with applications and services. On a broader scale, generative AI emerges as a formidable technological force, poised to reshape the landscape of Web 3.0. As generative AI continues its evolution and maturation, its prominence within the Web 3.0 ecosystem is bound to witness significant expansion.

\section{Conclusion}
\label{sec:Conclusion}
 Web 3.0 is an emerging technology that has the potential to bring a tremendous revolution in  various fields, such as finance, IoT, health informatics, education, distributed applications, and supply chain management. In this article, we have presented a comprehensive survey on how technologies can enable, empower, and revolutionize Web 3.0. Firstly, we have presented an overview of Web 3.0, discussed its effectiveness, and highlighted a number of applications and recent industry standards where the technologies were potentially deployed. Then, we have discussed and demonstrated various important technologies that will play key roles in the future development of Web 3.0, (i.e., IoT, 5G, blockchain, semantic web, and 3D interactive web technologies). For each technology, we have provided an overview, examined the state-of-the-art, and discussed how it can be utilized in future development of Web 3.0 scenarios. Finally, we have discussed open issues and some potential solutions that will pave the way for the wide adoption and deployment of Web 3.0 in the near future.


\end{document}